\documentclass[structabstract]{aa}

\usepackage{txfonts}
\usepackage{graphicx}
\usepackage{array}
\usepackage{natbib}
\bibpunct{(}{)}{;}{a}{}{,}

\begin{document}
%%%%%%%%%%%%%%%%%%%%%%%%%%%%%%%%%%%%%%%%%%%%%%%%%%%%%%%%%%%%%%%%%%%%%%%%%%%%%%%%%%%%%%%%%%%%%%%%%%%%%%%%%%%%
%	Title
%%%%%%%%%%%%%%%%%%%%%%%%%%%%%%%%%%%%%%%%%%%%%%%%%%%%%%%%%%%%%%%%%%%%%%%%%%%%%%%%%%%%%%%%%%%%%%%%%%%%%%%%%%%%
\title{Tracing the evolutionary stage of Bok globules: CCS and NH$_3$\thanks{Based on observations obtained with the 100-m telescope of the MPIfR (Max-Planck-Institut f\"ur Radioastronomie) at Effelsberg and the 64-m Parkes radio telescope.  The Parkes radio telescope is part of the Australia Telescope National Facility which is funded by the Commonwealth of Australia for operation as a National Facility managed by CSIRO. }}% not 

\author{ C.~Marka\inst{1}
\and K.~Schreyer\inst{1}
\and R.~Launhardt\inst{2}
\and D.~A.~Semenov\inst{2}
\and Th.~Henning\inst{2}}

\institute{Astrophysikalisches Institut und Universit\"atssternwarte (AIU), Schillerg\"a\ss chen 2-3, D-07745 Jena, Germany
\and Max-Planck-Institut f\"ur Astronomie (MPIA), K\"onigstuhl 17, D-69117 Heidelberg, Germany}	% addresses

\date{Received /
	Accepted }

\abstract {}
{This work pursues the investigation of a previously proposed correlation between chemical properties and physical evolutionary stage of isolated low-mass star-forming regions. In the past, the $N_\mathrm{NH_3}$/$N_\mathrm{CCS}$ abundance ratio was suggested to be a potentially useful indicator for the evolutionary stage of cloud cores. We aim to study its applicability for isolated Bok globules.}
{A sample of 42 Bok globules with and without signs of current star formation was searched for CCS($2_1$--$1_0$) emission, the observations were complemented with NH$_3$ measurements available in the literature and own observations. The abundance ratio of both molecules is discussed with respect to the evolutionary stage of the objects and in the context of chemical models.}
{The $N_\mathrm{NH_3}$/$N_\mathrm{CCS}$ ratio could be assessed for 18 Bok globules and is found to be moderately high and roughly similar across all evolutionary stages from starless and prestellar cores towards internally heated cores harbouring protostars of Class~0, Class~I or later. 
Bok globules with extremely high CCS abundance analogous to carbon-chain producing regions in dark cloud cores are not found. 
The observed range of $N_\mathrm{NH_3}$/$N_\mathrm{CCS}$ hints towards a relatively evolved chemical state of all observed Bok globules. 
}
{}

\maketitle

%%%%%%%%%%%%%%%%%%%%%%%%%%%%%%%%%%%%%%%%%%%%%%%%%%%%%%%%%%%%%%%%%%%%%%%%%%%%%%%%%%%%%%%%%%%%%%%%%%%%%%%%%%%%
\section{Introduction} 
\begin{table}
\caption{Overview of the observed globules.}
\label{table_overview}
\centering
\begin{tabular}{lllcc}
\hline \hline
Object		& RA 		& Dec 		& ~Associated	& Distance \\
		& (B1950)	& (B1950)	& ~IRAS Source	& (Ref.) \\
	& (\ h \ m \ s \ ) & (\ \degr \ \ \arcmin \ \ \arcsec \ \ )& & [pc] \\
\hline
CB\,3		& 00 25 59.0	&+56 25 32	& 00259+5625	& 2500~(2)\\
CB\,6		& 00 46 34.4	&+50 28 25	& 00465+5028	& 600 ~~(3)\\
CB\,12		& 01 35 24.3	&+64 47 59	& 01354+6447	& 800 ~~(2)\\
CB\,17		& 04 00 32.9	&+56 47 52	& (04005+5647)	& 250 ~~(3)\\
CB\,22		& 04 37 23.5	&+29 49 17	& \ldots 	& 140 ~~(2)\\
CB\,23		& 04 40 20.5	&+29 33 26	& \ldots 	& 140 ~~(2)\\
CB\,26		& 04 55 56.1	&+52 00 17	& 04559+5200	& 140 ~~(3)\\
CB\,28		& 05 03 51.3	&$-$04 00 18	& 05038-0400	& 450 ~~(3)\\
CB\,34		& 05 44 05.7	&+20 59 30	& 05440+2059	& 1500 (3)\\
CB\,44		& 06 04 43.0	&+19 28 19	& (06047+1923)	& 700 ~~(3)\\
CB\,68		& 16 54 27.2	&$-$16 04 48 	& 16544-1604	& 160 ~~(3)\\
CB\,125		& 18 12 39.0	&$-$18 12 14	& (18121-1813)	& 200 ~~(2)\\
CB\,179		& 19 01 57.0	&$-$05 25 35	& (19018-0525)	& 200 ~~(2)\\
CB\,188		& 19 17 54.0	&+11 30 10	& 19179+1129	& 300 ~~(3)\\
CB\,222		& 20 32 49.9	&+63 52 00	& 20328+6351	& 400 ~~(3)\\
CB\,224		& 20 35 30.9	&+63 42 48	& 20355+6343	& 400 ~~(3)\\
CB\,230		& 21 16 50.8	&+68 04 52	& 21169+6804	& 400 ~~(3)\\
CB\,232		& 21 35 14.4	&+43 07 17	& 21352+4307	& 600 ~~(3)\\
CB\,243		& 23 22 51.9	&+63 20 11	& 23228+6320	& 700 ~~(3)\\
CB\,244		& 23 23 48.9	&+74 01 10	& 23238+7401	& 200 ~~(3)\\
CB\,246		& 23 54 12.0	&+58 17 27	& \ldots 	& 140 ~~(3)\\
BHR\,12		& 08 07 39.0	& $-$35 55 54	&  08076-3556	& 400 ~~(1)\\
BHR\,13		& 08 26 44.1	& $-$33 36 31	&  08267-3336	& 400 ~~(1)\\
BHR\,15		& 07 14 27.5	& $-$43 52 26	&  07144-4352	& 400 ~~(1)\\
BHR\,23-1	& 08 33 42.6	& $-$40 28 02	&  08337-4028	& 500 ~~(1)\\
BHR\,28		& 07 26 20.0	& $-$50 58 18	&  \ldots	& 400 ~~(1)\\
BHR\,34		& 08 25 03.4	& $-$50 30 34	&  08250-5030	& 200 ~~(3)\\
BHR\,36		& 08 24 15.9	& $-$50 50 44	&  08242-5050	& 400 ~~(1)\\
BHR\,41		& 08 26 11.5	& $-$51 00 39	&  08261-5100	& 400 ~~(1)\\
BHR\,55		& 09 44 57.0	& $-$50 52 06	&  09449-5052	& 300 ~~(1)\\
BHR\,59		& 11 05 03.0	& $-$61 49 36	&  \ldots 	& 200 ~~(4)\\
BHR\,71		& 11 59 03.1	& $-$64 52 11	&  11590-6452	& 200 ~~(3)\\
BHR\,74		& 12 19 21.0	& $-$66 10 30	&  \ldots 	& 175 ~~(4)\\
BHR\,86		& 13 03 41.4	& $-$76 44 03	&  13036-7644	& 180 ~~(3)\\
BHR\,111	& 15 38 34.0	& $-$52 38 21	&  \ldots 	& 250 ~~(4)\\
BHR\,121-1	& 16 54 59.6	& $-$50 30 58	&  16549-5030	& 125 ~~(4)\\
BHR\,121-2	& 16 55 27.8	& $-$50 31 01	&  16554-5031	& 125 ~~(4)\\
BHR\,137	& 17 18 08.6	& $-$44 06 17	&  (17181-4405)	& 700 ~~(3)\\
BHR\,138	& 17 15 54.0	& $-$43 24 04	&  17159-4324	& 225 ~~(4)\\
BHR\,139-1	& 17 17 15.9	& $-$43 16 54	&  17172-4316	& 225 ~~(4)\\
BHR\,140-1	& 17 19 18.8	& $-$43 19 24	&  17193-4319	& 225 ~~(4)\\
BHR\,148	& 17 01 09.2	& $-$36 13 59	&  17011-3613	& 175 ~~(4)\\
\hline
\end{tabular}
\tablebib{(1)~\citet{Bourke95b}; (2)~\citet{LH97}; (3)~\citet{L2010} and private communication; (4)~\citet{Racca09}.}
\end{table}

Bok globules, named in honor of the astronomer Bart Bok who drew attention to those objects and their possible role in the star formation process \citep{BR47}, appear as small and isolated dark clouds. Although the majority of stars in the Galaxy is formed in Giant Molecular Cloud complexes, the small globules have been recognized as particularly interesting targets for a study of low-mass star formation, since they represent a less complex environment and are therefore more easily described by theoretical models. 
The conditions inside Bok globules and the properties of the young stellar objects (YSOs) embedded therein have been studied at various wavelengths  in several surveys in the past. Thermal radio emission arising from dust is observed at centimeter \citep{Moreira97, Moreira99} and millimeter wavelengths (\citealt{LH97}, hereafter \defcitealias{LH97}{LH97}LH97; \citealt{LH98c}); submillimeter emission gives hints for deeply embedded protostars \citep{Huard99}. With near-infrared observations, candidates for YSOs are identified and examined \citep{Yun95, Alves94, Racca09}. Molecular line observations provide information about the physical conditions, e.g. temperature, density, and magnetic fields, and trace the velocity structure of the clouds. They can be used to identify collapsing clouds \citep{Wang95} and reveal the presence of powerful molecular outflows \citep{Yun92,Yun94a}.
A comprehensive overview of star formation in different stages is achieved by multi-wavelength studies (\citealt{L2010}, hereafter\defcitealias{L2010}{L2010}L2010).

In globules and cloud cores a large variety of molecules has been identified. The complex chemical processes leading to their formation and destruction are ruled by the local conditions, including surface reactions on ice-covered grains in the coldest and densest parts of the cloud, and are influenced by the heating of a forming star within the core or the ionizing interstellar radiation close to the cloud surface \citep[review by,  e.g.,][]{Dishoeck98}. While ammonia was the first polyatomic molecule to be discovered in the interstellar medium \citep[review e.g. by][]{Townes83}, the CCS radical has been identified only two decades ago \citep{Saito87} and the main pathways leading to its formation are still under discussion \citep{Suzuki92,Petrie96,Sakai07}, although it is frequently observed along with other carbon-chain molecules \citep[e.g.,][]{Hirota2009}.

From a study of CCS and NH$_3$ among other molecules, \citet{Suzuki92} found that carbon-chain molecules are abundant in the early evolutionary phases of dark cloud cores, while ammonia tends to be more abundant in more evolved, actively star-forming regions. This result was also supported by their chemical model calculations, where CCS is formed but also destroyed rapidly, while in turn a replenishment of the molecule is impeded by the increasing lock-up of carbon atoms in CO. In contrast, NH$_3$ forms in a sequence of very slow reactions and reaches its highest abundance in the late evolutionary phases. Therefore they proposed the $N_\mathrm{NH_3}$/$N_\mathrm{CCS}$ abundance ratio to be a possibly useful indicator for the evolutionary stage of star-forming clouds.
In the following years, the abundance of both molecules in a number of Bok globules and around Herbig Ae/Be stars \citep{SC96}, as well as in a larger sample of intermediate-mass and low-mass star-forming regions with H$_2$O maser emission \citep[][hereafter dGM06]{dGM06}\defcitealias{dGM06}{dGM06}was examined. However, both studies suffered from a small number of CCS detections, which hindered a definitive judgement about this hypothesis.
Focusing on statistics on a large number of dense cores within the Perseus Molecular cloud, a recent work by \citet{F09} provided evidence for a lower fractional abundance of CCS in protostellar cores compared to starless cores.

In this work we are focusing on small Bok globules, typically forming only single or few low-mass stars, since they represent the most secluded star-forming environments available for investigation. 
An interference by other nearby, possibly high-mass, forming stars can be excluded. Moreover, a mixing with ambient material which could modify the chemical composition and exacerbate a comparison with theoretical models, as it is possible in the case of a dense core inside a larger cloud complex, is implausible for the isolated globules. An overview of the sample examined in this work is given in Sect.~\ref{sample}, followed by a description of the observational results and analysis in Sect.~\ref{results}. The discussion of the derived physical parameters is presented in Sect.~\ref{discussion}, concluding remarks are given in Sect.~\ref{summary}.

%%%%%%%%%%%%%%%%%%%%%%%%%%%%%%%%%%%%%%%%%%%%%%%%%%%%%%%%%%%%%%%%%%%%%%%%%%%%%%%%%%%%%%%%%%%%%%%%%%%%%%%%%%%%
\section{Description of the sample} 
\label{sample}
Our sample consists of 42 Bok globules from the catalogs of \citet{CB88} for the northern hemisphere and \citet{Bourke95a} for the southern globules. 
The northern sources have been searched for NH$_3$(1,1) and (2,2) emission with the Effelsberg 100-m telescope in the survey of \citet[hereafter L96]{L96},\defcitealias{L96}{L96} while the southern globules are included in the NH$_3$ survey of \citet[hereafter BHR95b]{Bourke95b},\defcitealias{Bourke95b}{BHR95b}so that a uniform set of ammonia measurements is available for comparison with our data. Table~\ref{table_overview} lists the positions of the CCS measurement, the IRAS point sources associated with the globules (sources in brackets were located outside the telescope beam), and their distance.
Following the classification scheme of \citetalias{L2010}, we sorted the objects into three groups according to the presence and evolutionary class \citep[cf.][]{Lada87,Andre93} of embedded YSOs: 
\begin{itemize}
 \item Group $-$I: Starless or prestellar globules, or only very young \hspace*{1.4cm} and very low-luminosity source embedded.
 \item Group ~\,0: Globules harbouring a Class~0 protostar.
 \item Group ~~I: Globules containing YSOs of Class~I or later.
\end{itemize}
The classification of the individual clouds is indicated in Table~\ref{table_results}. Where too sparse observations of the specific globule prevent a reliable identification of the evolutionary group, a presumptive classification is given in brackets. It should be emphasized that especially prestellar cores or Class~0 protostars will most probably remain unidentified when there is a lack of detailed maps of molecular emission and/or millimeter continuum emission.
A detailed discussion for most globules of the sample can be found in \citetalias{L2010} and the references therein, while a number of southern globules has been classified in \citet{Racca09}. In the following we comment therefore only shortly on the globules not described there.

\textit{CB\,3}. A compact submillimeter source without infrared counterpart is located at the center of a bipolar molecular outflow \citep{Yun94a} and was suggested to consist of an aggregate of Class~0 sources by \citet{Huard2000}. The source IRAS\,00259+5625 is located 15\arcsec\, away, its near-infrared counterpart was classified as Class~II object \citep{Yun94b, Yun95}. The globule also harbours a H$_2$O-maser (\citealt{Scappini91},\citetalias{dGM06}); it is believed to be rather an evolved intermediate to high-mass star-forming region \citep{LH98b,CB99}.

\textit{CB\,12}. An associated IRAS point source is detected at 60 and 100\,$\mu$m, but neither a molecular outflow \citep{Yun92} nor dust continuum emission at 1.3\,mm \citepalias{LH97} point towards the presence of an embedded protostar.

\textit{CB\,22 and CB\,23}. Both globules are not detected in the far-infrared (no IRAS sources) or in the 1.3\,mm dust continuum \citepalias{LH97}. According to the shape of spectral lines, CB\,22 seems to be quiescent while CB\,23 shows signatures of a possible infall motion \citep{Lee99,Lee04}. 

\textit{CB\,28}. Only emission longward of 60\,$\mu$m was detected by IRAS towards this globule; searches for outflows \citep{Yun94a} and YSO candidates in the near-infrared \citep{Yun94b} resulted in no detections. 

\textit{CB\,34}. Numerous studies suggest multiple star formation in the globule. Five submillimeter sources were detected by \citet{Huard2000} and classified as probable Class~0 objects. The aggregate of protostars is associated with a system of multiple outflows \citep{Codella03} and jets \citep{Moreira95}. Evidence for the presence of more evolved Class~I and Class~II objects is provided by the detection of several very red objects in the near-infrared \citep{Alves95}.

\textit{CB\,44}. The associated IRAS source cited by \citet{CB88} resides at the very edge of the globule. Two 3.6\,cm continuum sources without infrared counterparts, candidate Class~0 objects, were detected by \citet{Moreira99} but no detection was possible at shorter wavelengths (e.g. at 1.3\,mm by \citetalias{LH97}). 

\textit{CB\,125}. Several IRAS point sources are located in the vicinity, but none within the area of highest extinction visible on optical images of the globule. Both IRAS\,18127-1803 and IRAS\,18122-1818 have reliable fluxes only at 12 and 25\,$\mu$m, they were considered as candidate pre-main sequence star \citepalias{LH97} and YSO \citep{LeeMyers99}, respectively. The 100\,$\mu$m detection towards IRAS\,18130-1824 was considered as cirrus emission due to the lack of submillimeter continuum emission \citep{Huard99}. Line observations in CS towards IRAS\,18126-1820 at the southern edge of CB\,125 show unsuspicious gaussian line profiles \citep{Launhardt98}. Due to the the presence of candidate YSOs close to the cloud, we assume an evolved stage.

\textit{CB\,179}. The association with a cold infrared source detected only at 60 and 100\,$\mu$m, the absence of 1.3\,mm continuum emission \citepalias{LH97} and narrow, weak CO lines \citep{CYH91} suggest an early evolutionary stage.

\textit{CB\,222}. Neither is the associated IRAS source detected at wavelengths shortward of 60\,$\mu$m, nor is 1.3\,mm dust emission found towards this globule \citepalias{LH97}. Together with unsuspicious gaussian line profiles measured in CS \citep{Launhardt98}, we assume an early evolutionary stage for CB\,222.

\textit{BHR\,13}. The narrow CO line profiles observed towards this cometary globule are supposed to arise from cold, quiescent gas \citep{Otrupcek00}, while the associated IRAS source was identified as a T~Tauri star \citep{Sahu92}.

\textit{BHR\,15}. The IRAS source associated with this cometary globule is detected only at 60 and 100\,$\mu$m, and CO line profiles typical of cold, quiescent gas have been observed \citep{Otrupcek00}. Therefore, an early evolutionary stage is assumed.

\textit{BHR\,23}. \citet{Santos98} conclude from near-infrared observations that an aggregate of several YSOs might be harboured by this globule. High-velocity wings in CO lines detected by \citet{Ur2009} and \citet{Otrupcek00} might indicate the presence of a molecular outflow. Methanol maser emission, which is believed to be associated with high-mass star formation, was also observed towards this source \citep{Walsh97}.

\textit{BHR\,28, BHR\,59, BHR\,74 and BHR\,111}. These globules do not harbour any IRAS sources. Profiles of CO lines observed towards BHR\,74 and the cometary globule BHR\,28 suggest the presence of cold and quiescent gas \citep{Otrupcek00}. In BHR\,59 and BHR\,111 the detection of line wings might indicate the presence of several blended velocity components or outflow motions \citep{Otrupcek00}. However, no more detailed observations are available for those globules.

\textit{BHR\,41}. \citet{Santos98} classified the two near-infrared sources separated by 4\arcsec\, seen at the position of the IRAS source as likely Class~I objects. However, no 1.3\,mm dust emission was detected towards this source \citepalias{L2010}.

\textit{BHR\,137}. \citet{Santos98} detected three infrared sources at the IRAS position, two of which are likely Class~II objects. However, the IRAS source is located at the rim of the globule and it is unsure if is related. Emission in the 1.3\,mm dust continuum has been detected by \citet{LH98c}, as well as a blue wing in the CO emission. \citetalias{L2010} suggested a rather early evolutionary stage.\\

Some of the globules are known to contain multiple sources in different evolutionary stages, namely CB\,3 \citep{Yun94b,Huard2000}, CB\,34, 
CB\,224, CB\,232, CB\,243, CB\,244 and BHR\,12 \citepalias{L2010}. 
In such cases, where there are sources of different evolutionary stage within the beam of the NH$_3$ and CCS observations, we assigned the globule to the evolutionary group that agrees with the source from which most of the mm dust emission arises, i.e., which has the largest reservoir of cold gas and is therefore dominating the detected line emission. However, it cannot be excluded that one of the other sources also contributes to the detected line emission. Also, the maturity of individual globules might be underestimated, because the presence of an evolved YSO nearby a pre/protostellar core possibly affects its chemistry, especially since in most cases the projected separation between the differently 
evolved sources is only a few thousand AU. We have noted these cases in Table~\ref{table_results}.

% -----------------------------------------------------------------------------------------------
\begin{figure*}
\resizebox{\hsize}{!}{$\begin{array}{ccccccc}
\includegraphics[angle=270]{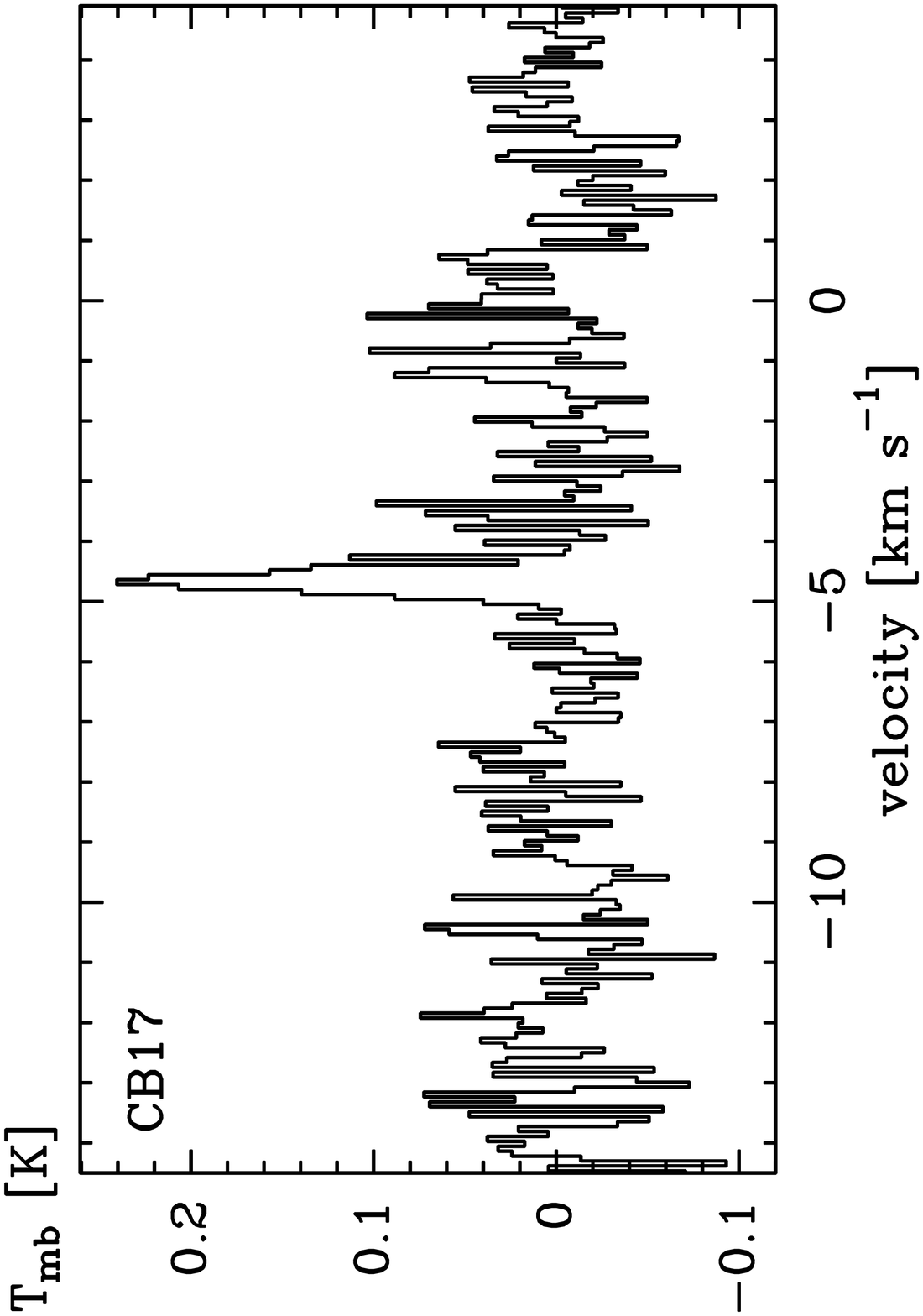}& &\includegraphics[angle=270]{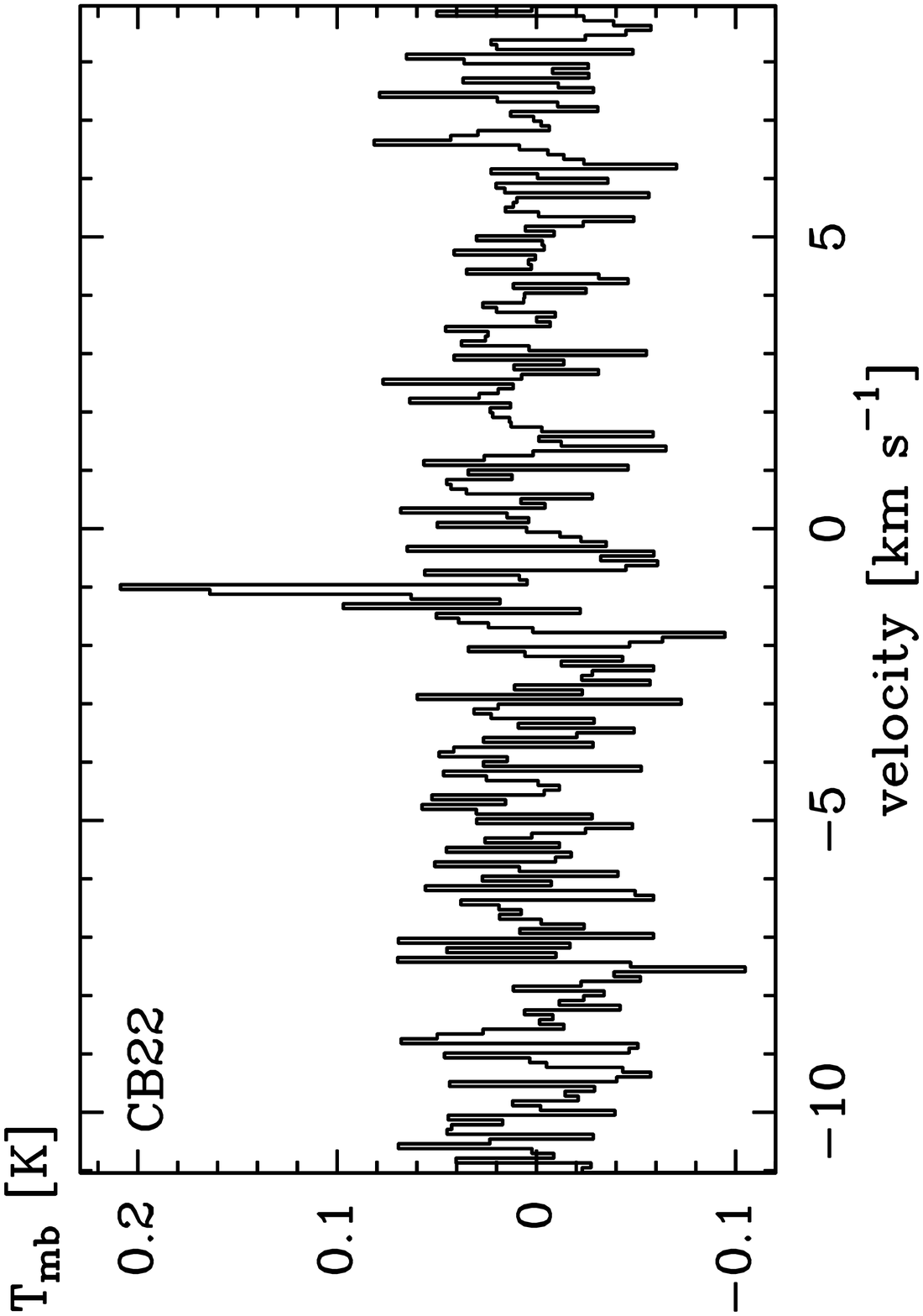}& &\includegraphics[angle=270]{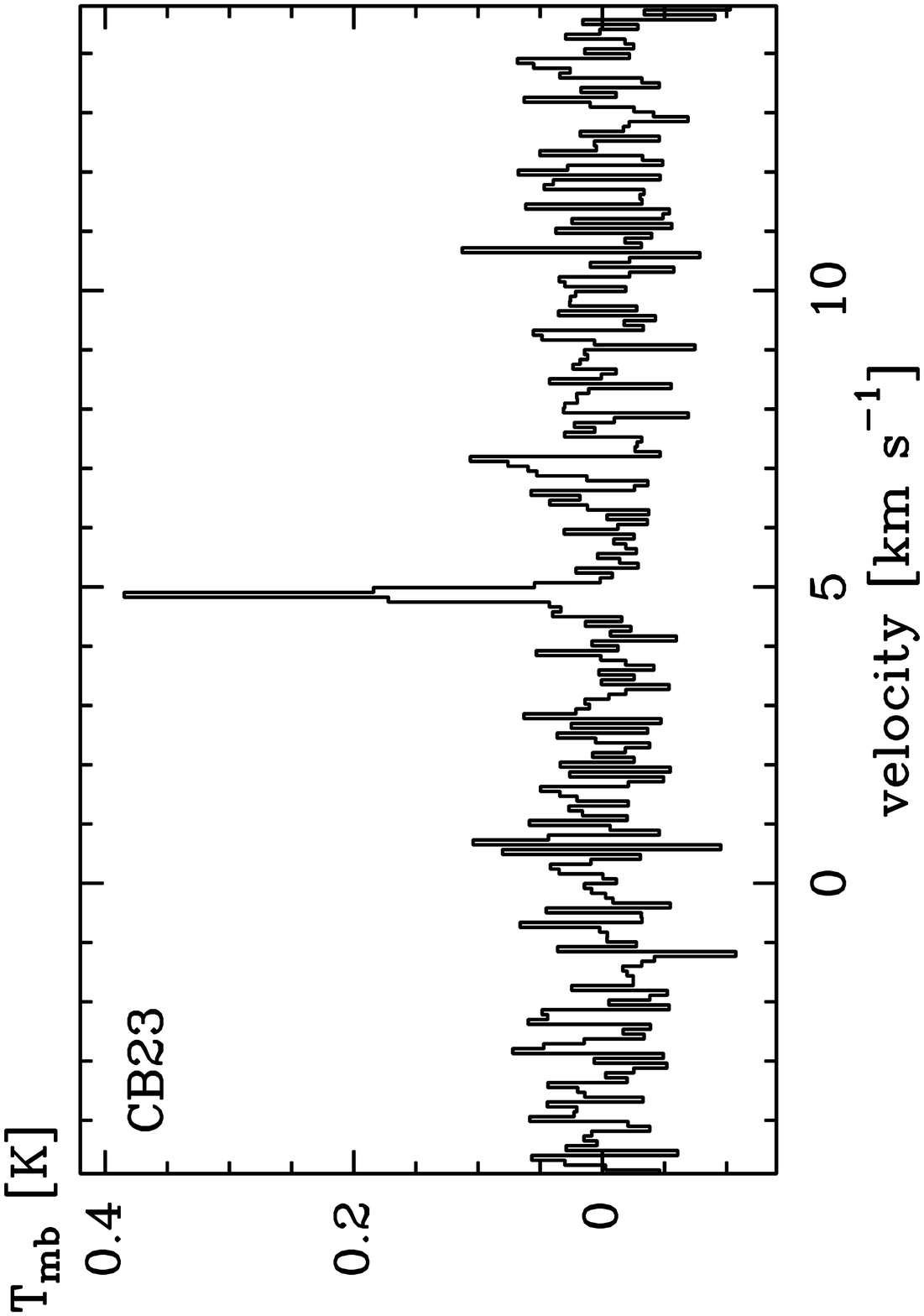}& &\includegraphics[angle=270]{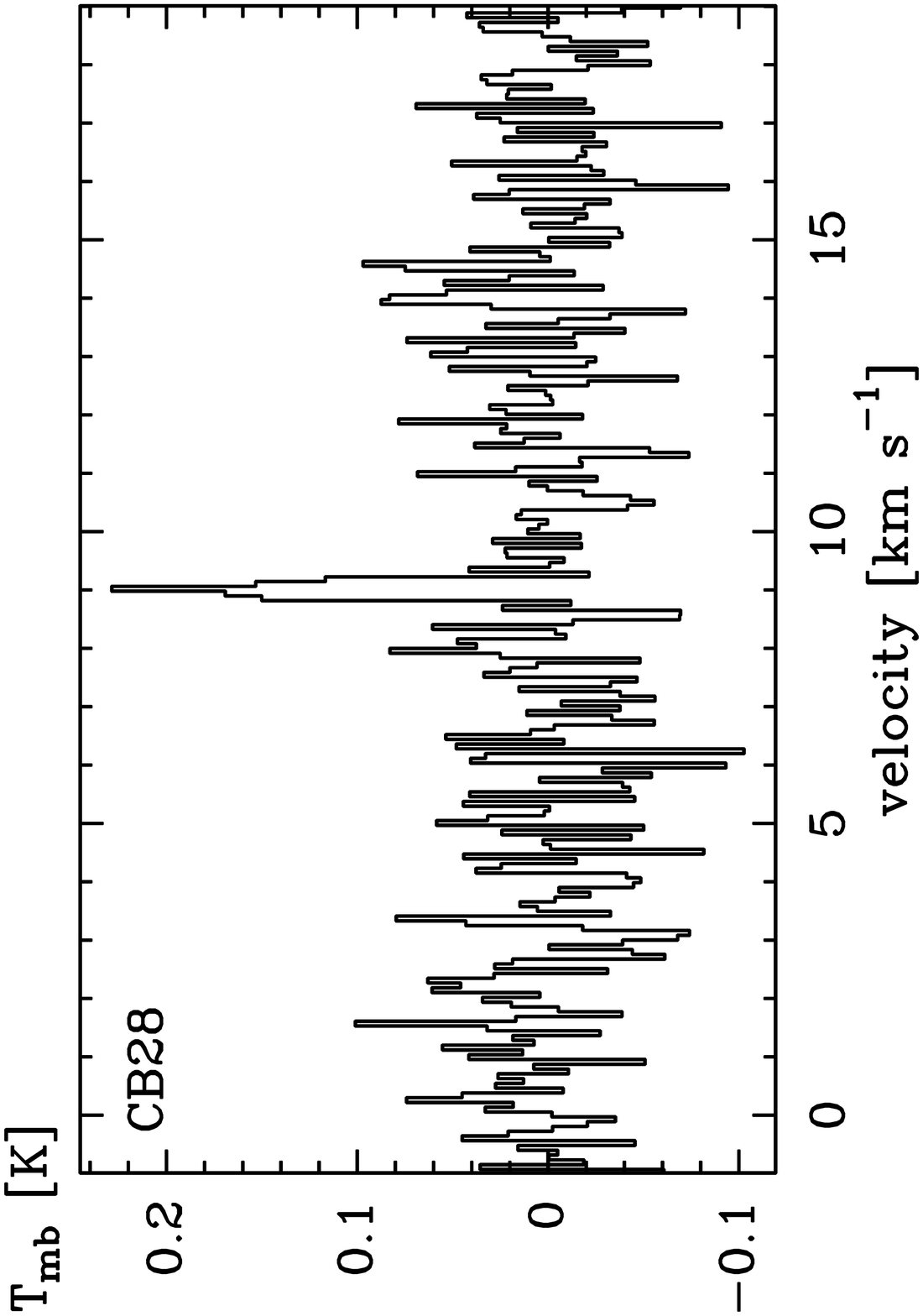}\\
\includegraphics[angle=270]{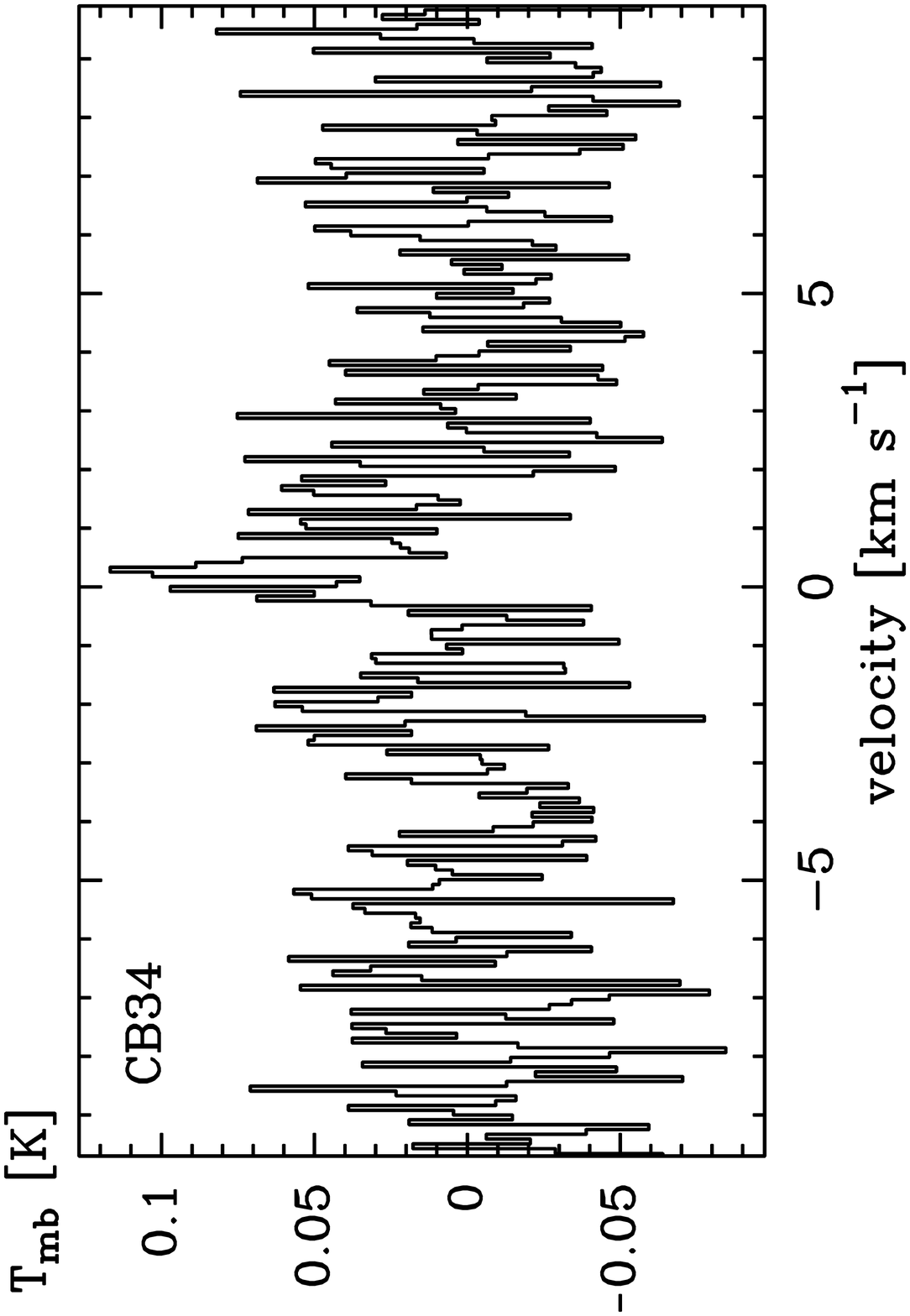}& & \includegraphics[angle=270]{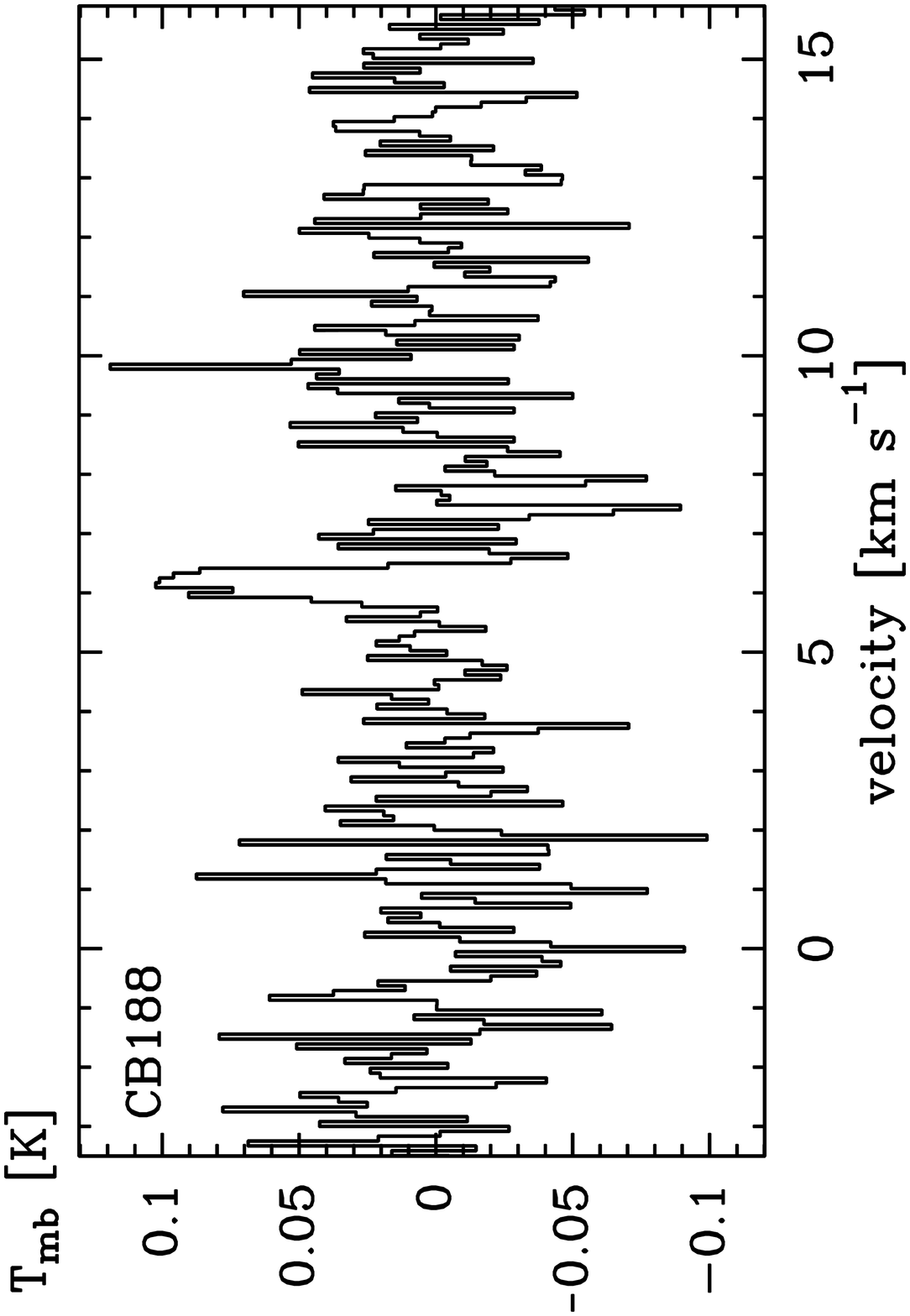}& &\includegraphics[angle=270]{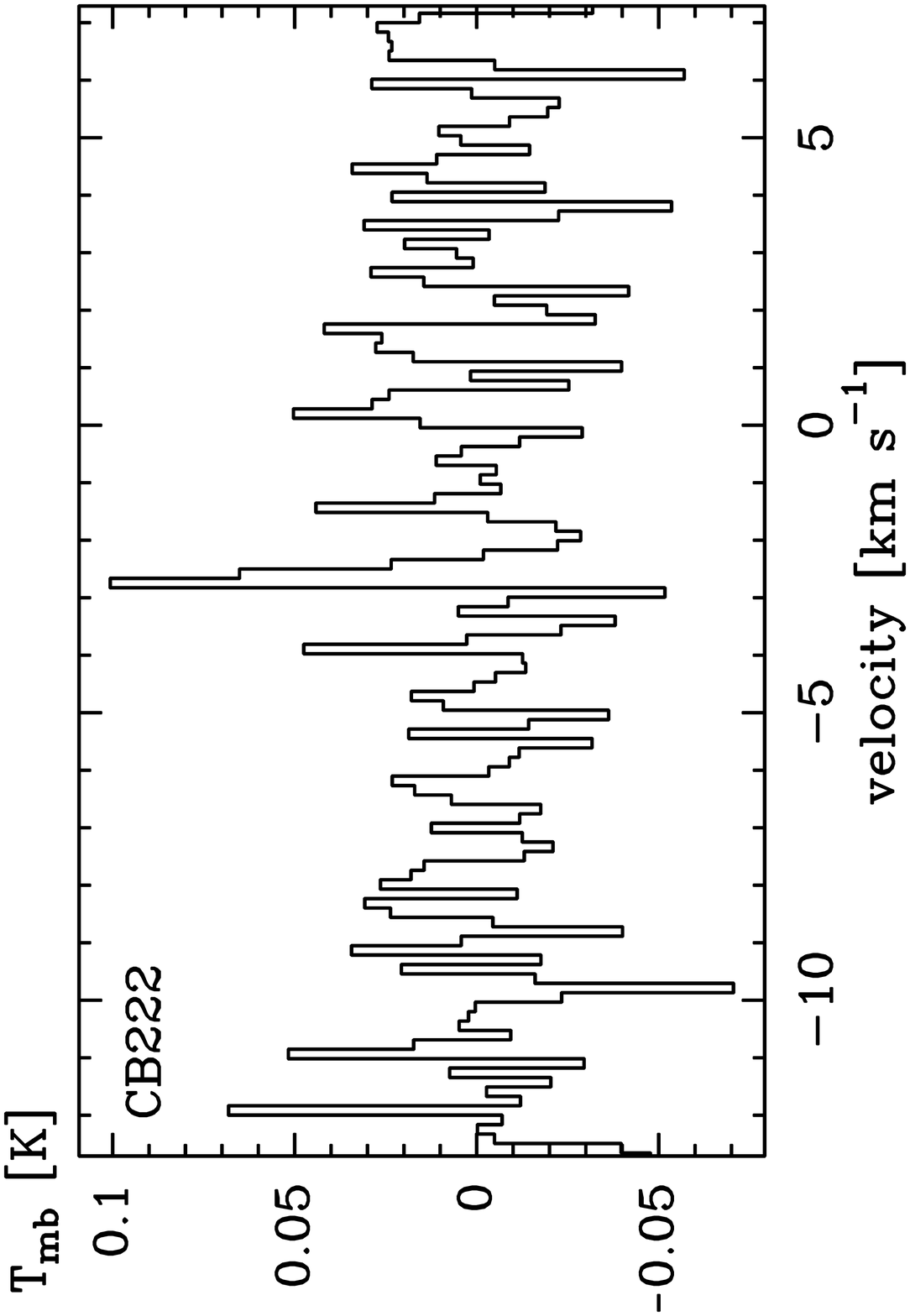}& &\includegraphics[angle=270]{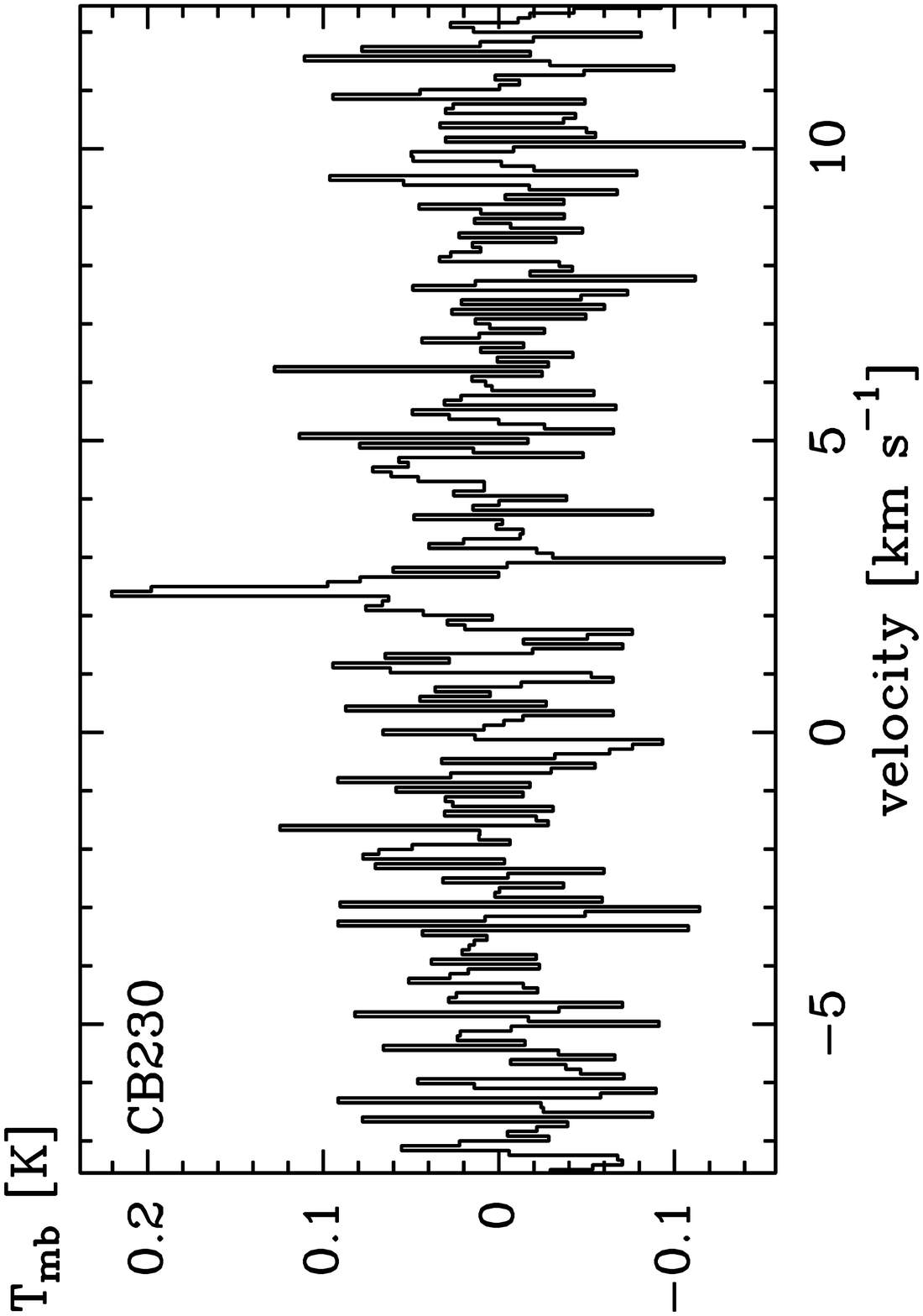}\\
\includegraphics[angle=270]{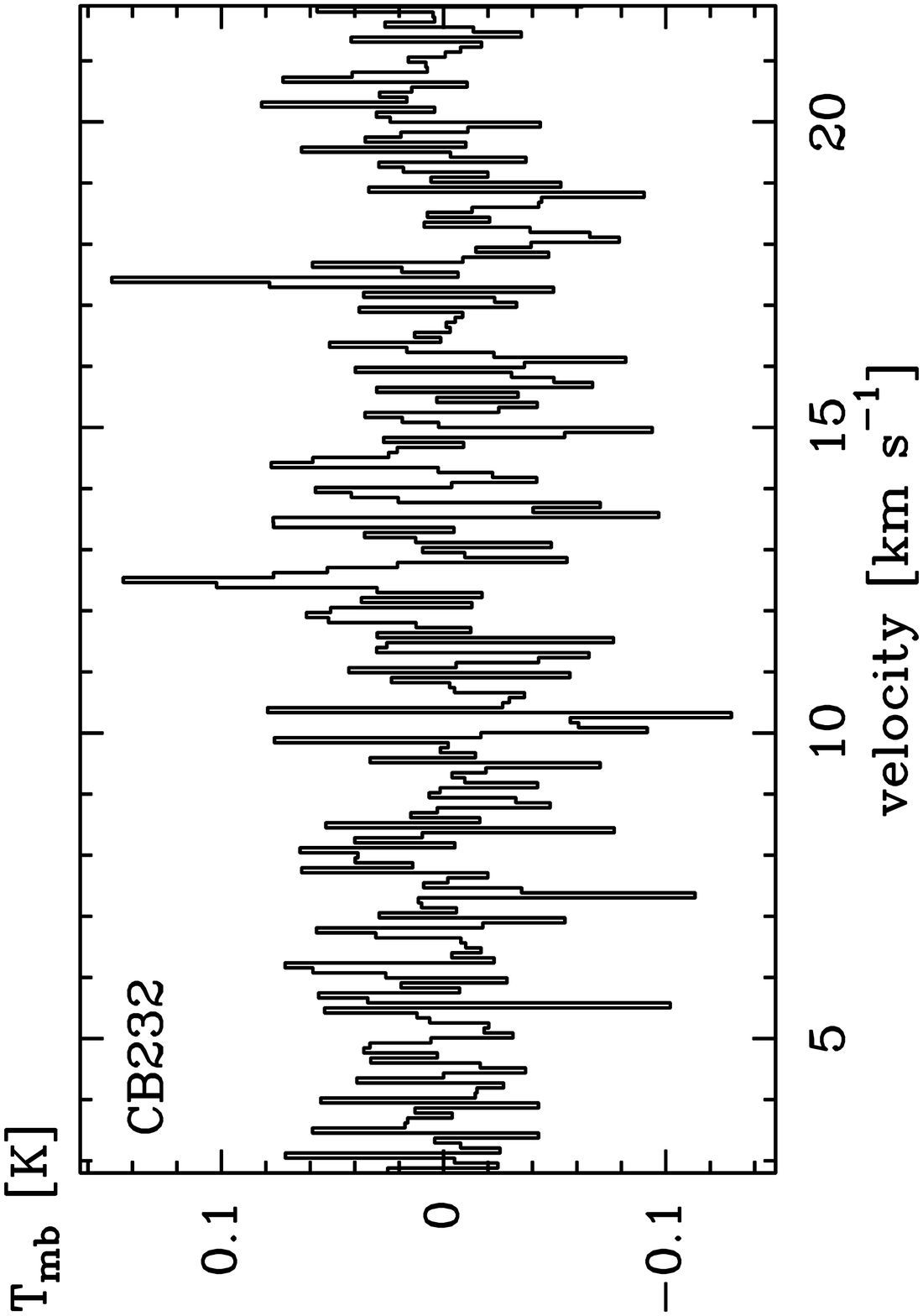}& &\includegraphics[angle=270]{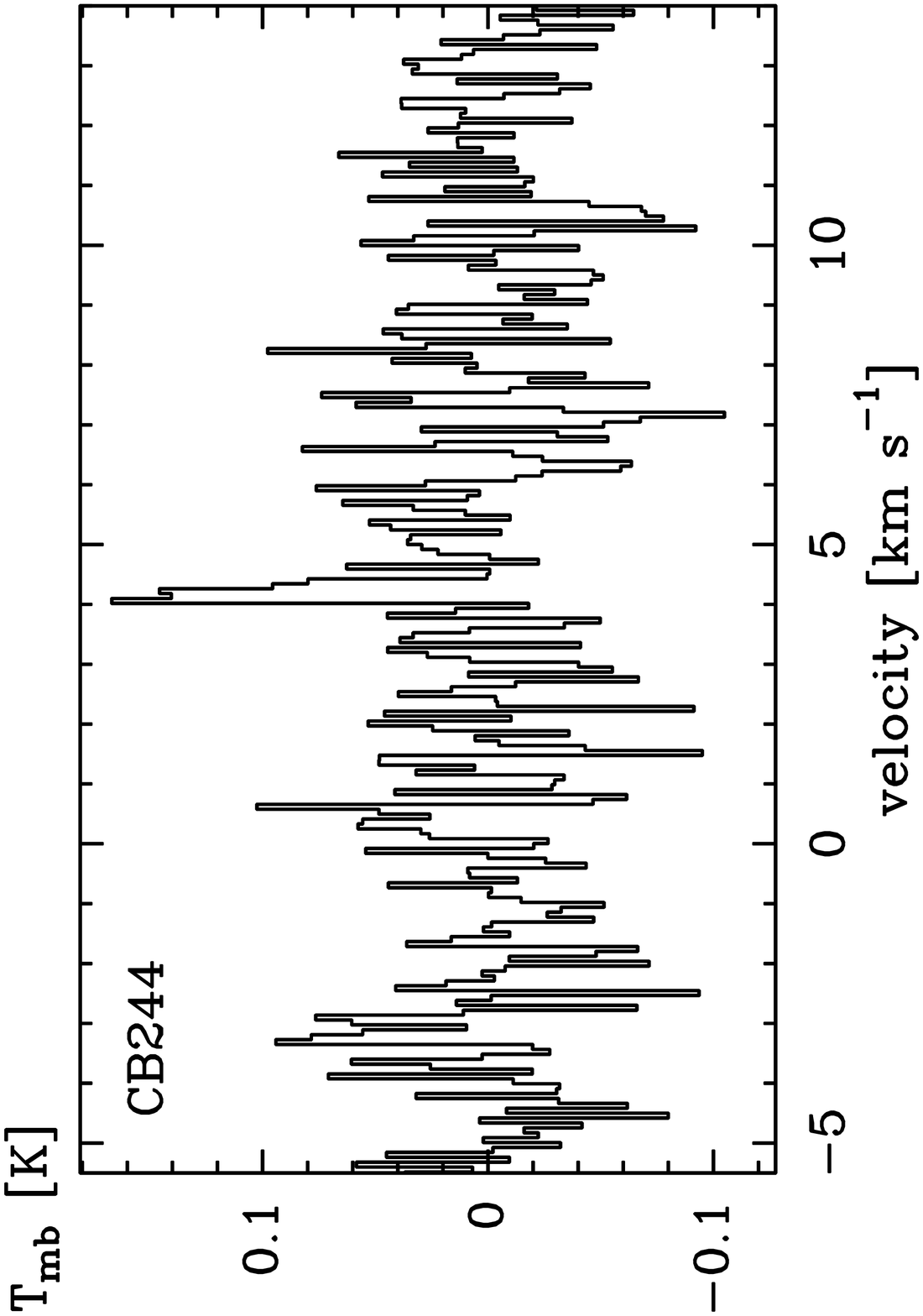}& &\includegraphics[angle=270]{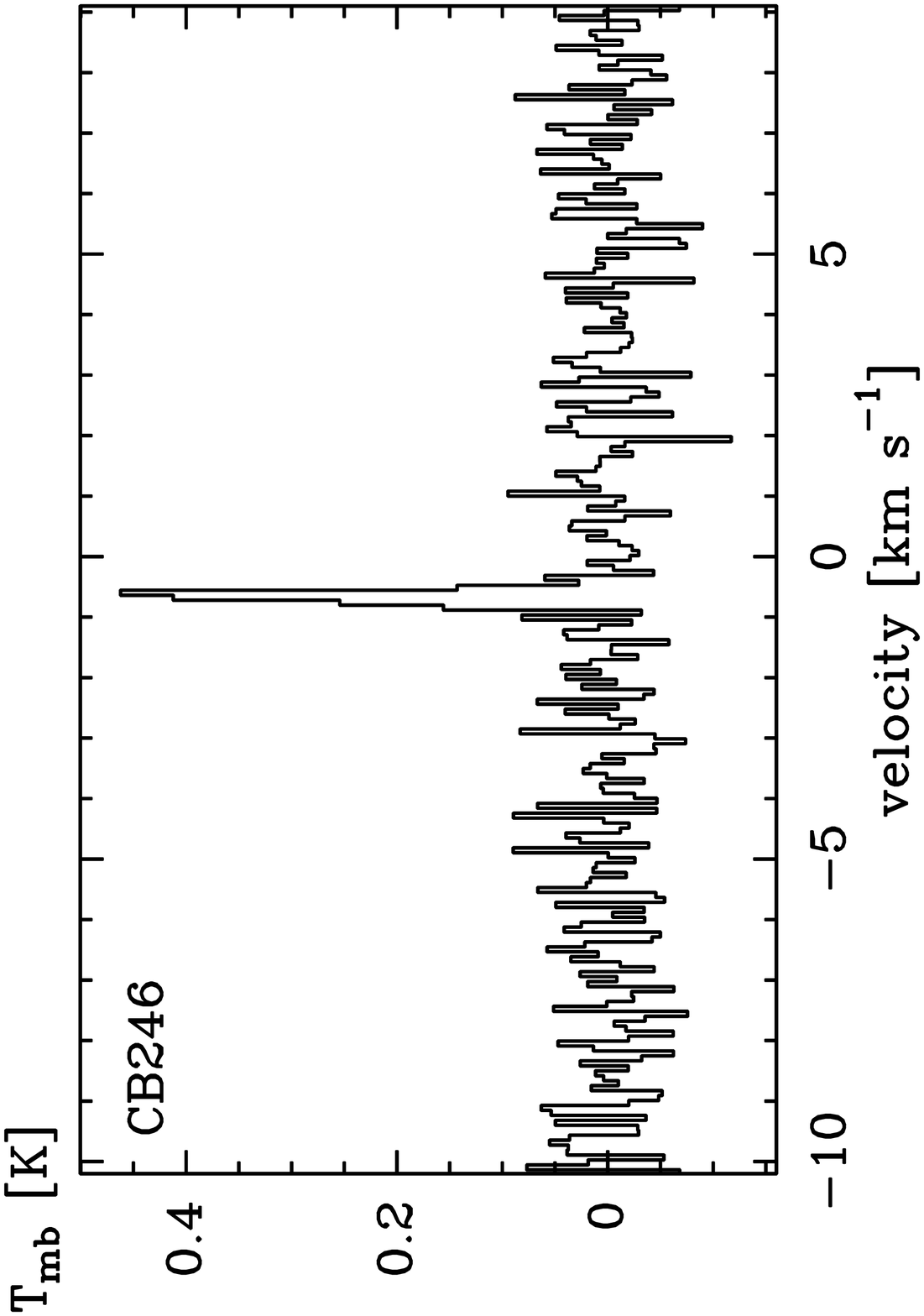}& &\includegraphics[angle=270]{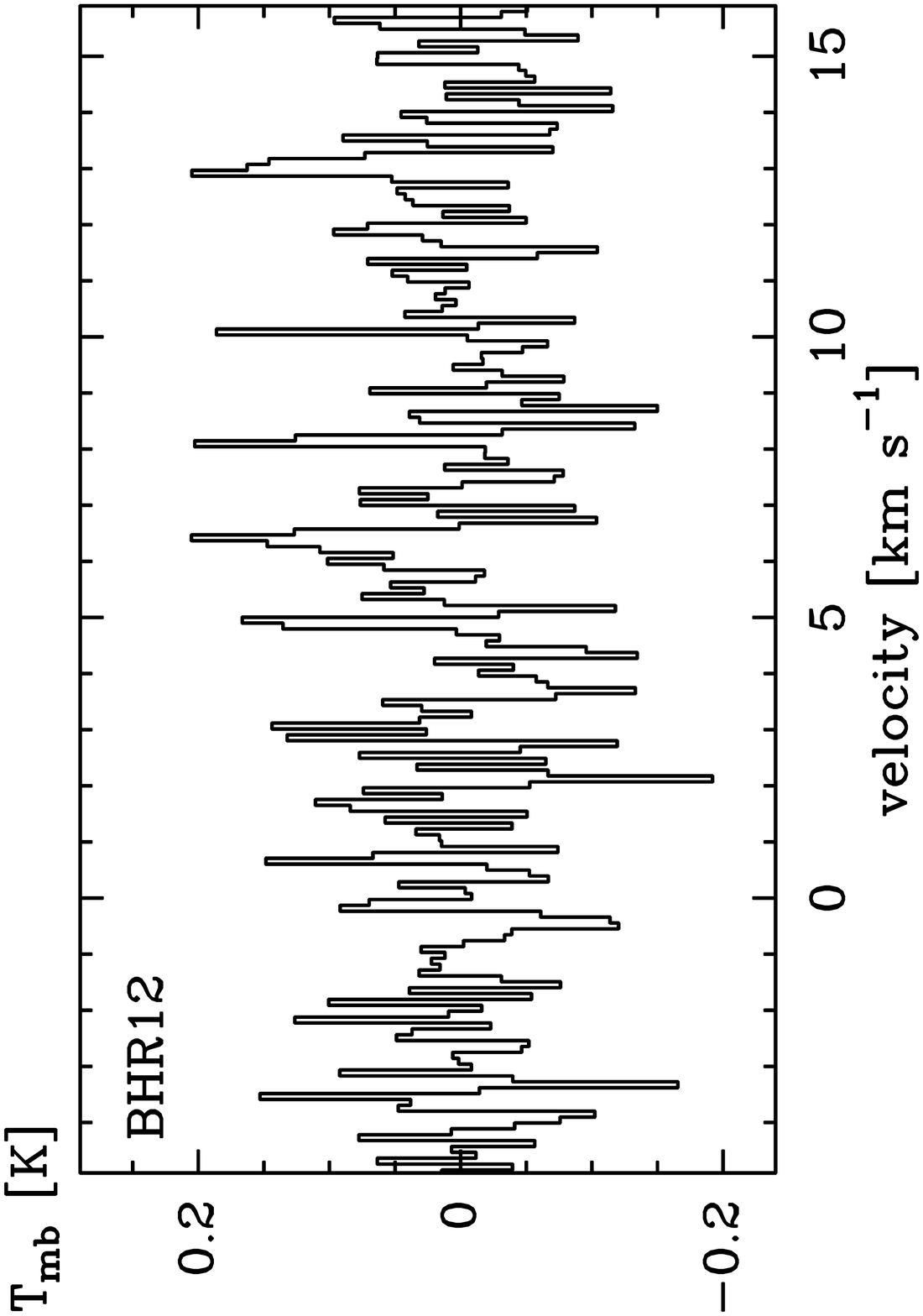}\\
\includegraphics[angle=270]{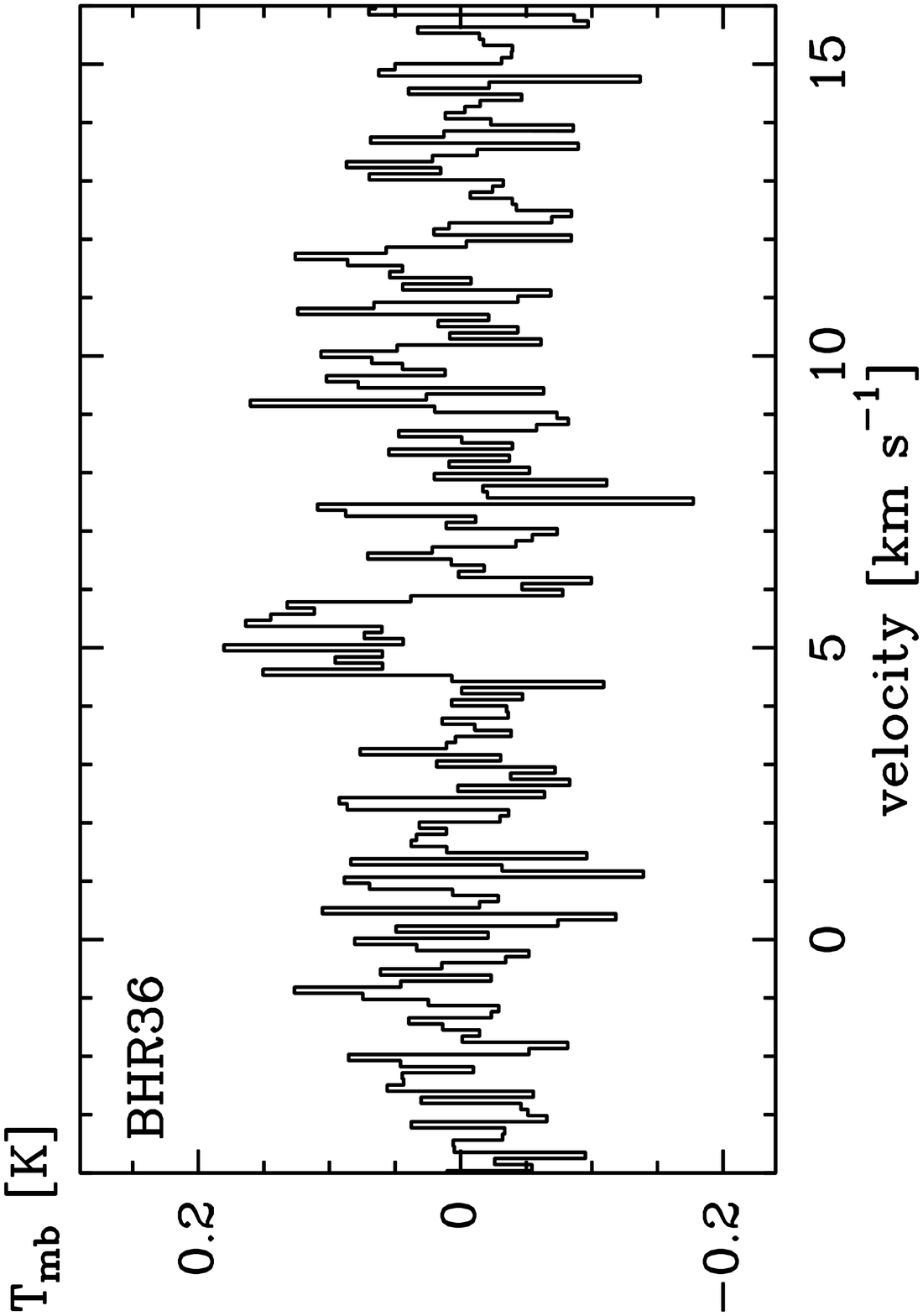}& &\includegraphics[angle=270]{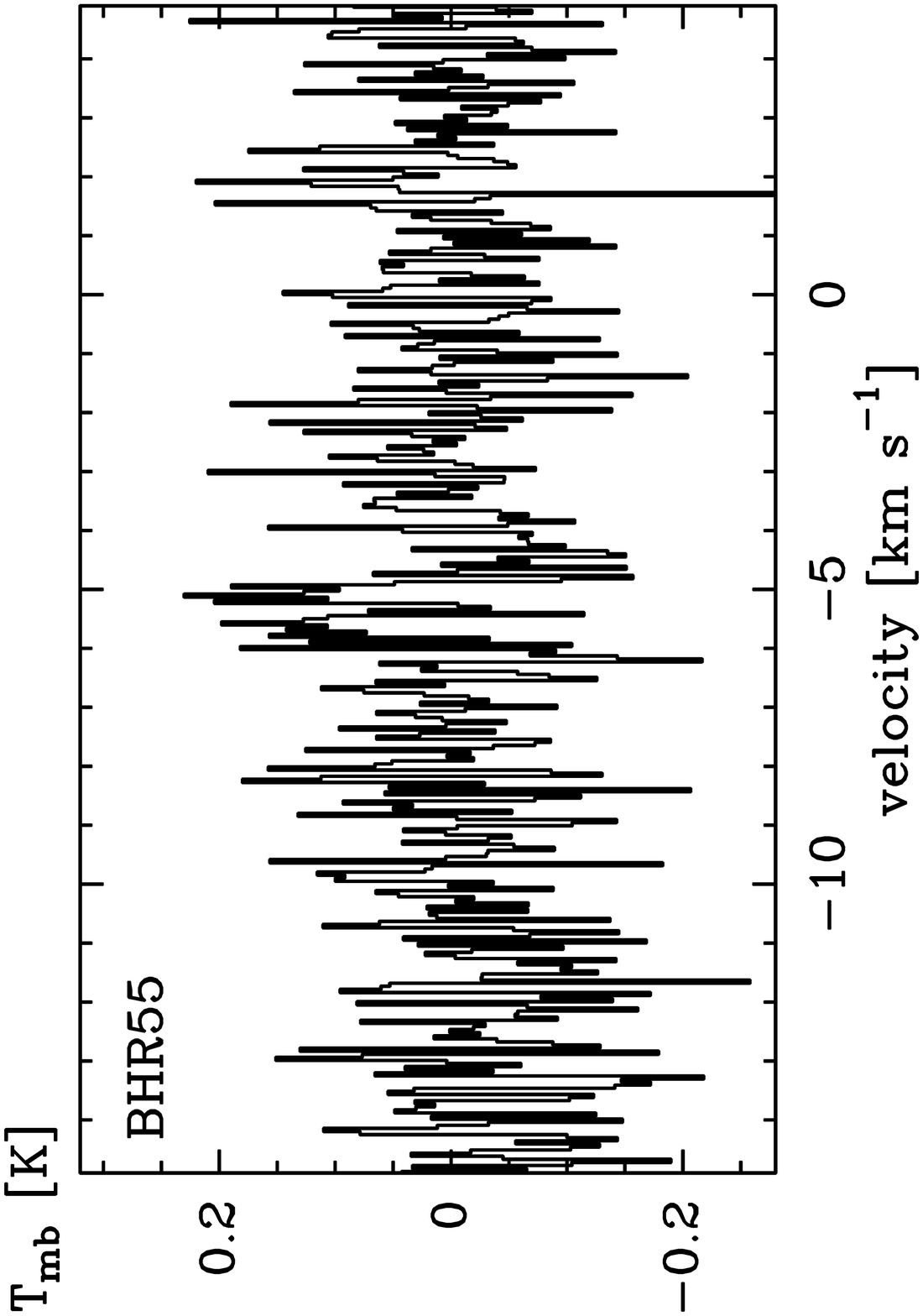}& &\includegraphics[angle=270]{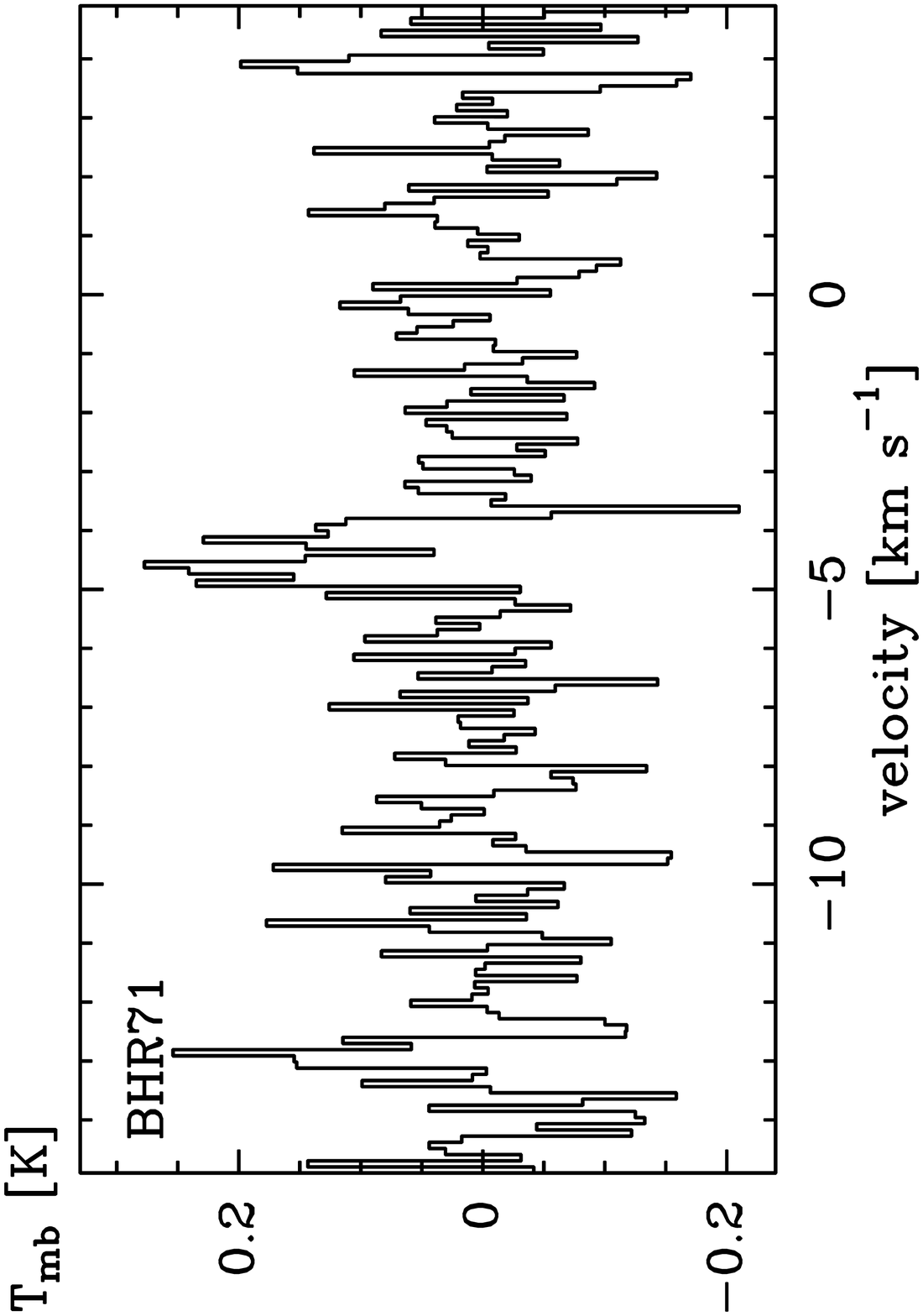}& &\includegraphics[angle=270]{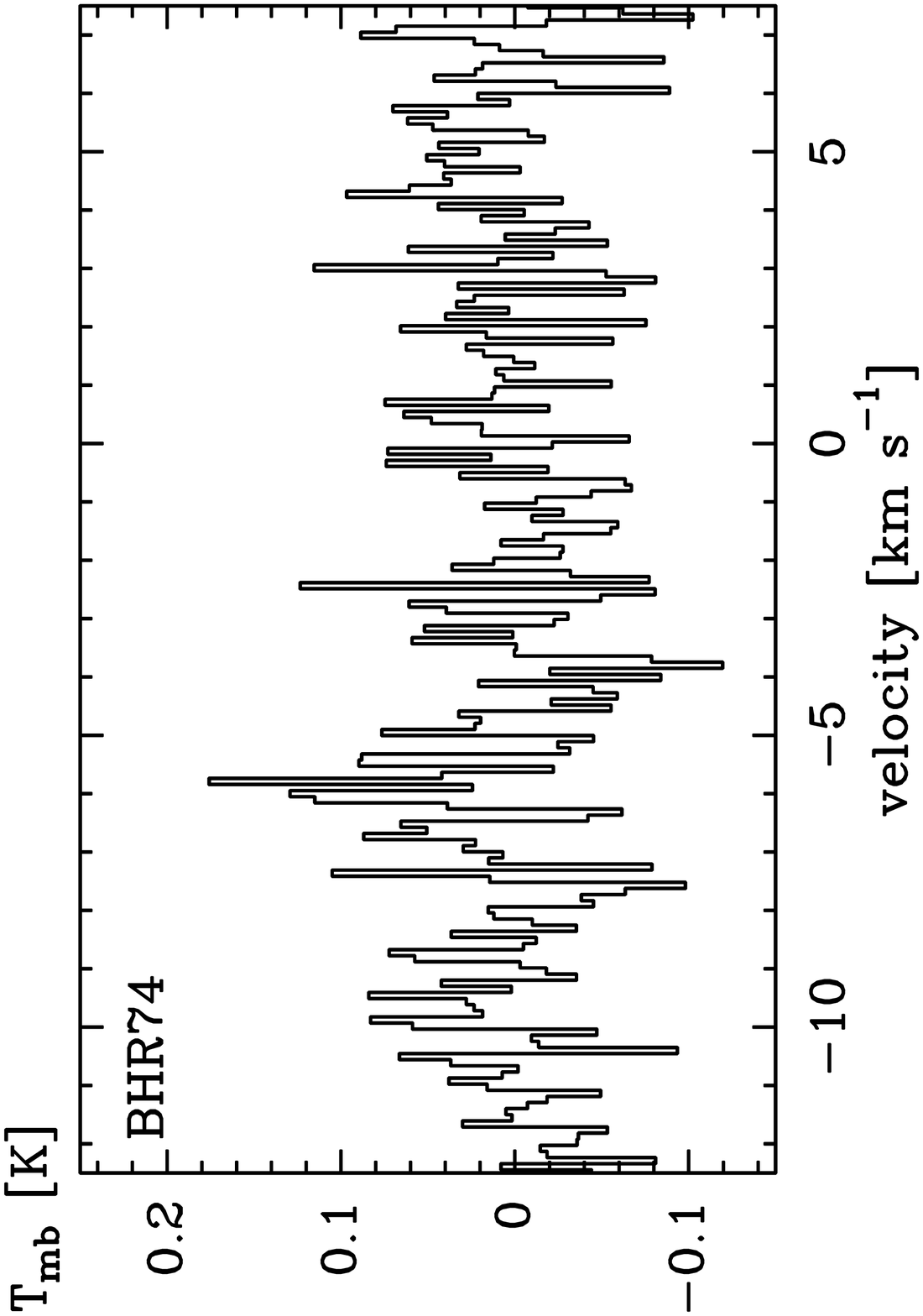}\\
\end{array}$}
\parbox{0.5\hsize}{\resizebox{\hsize}{!}{\includegraphics[angle=270]{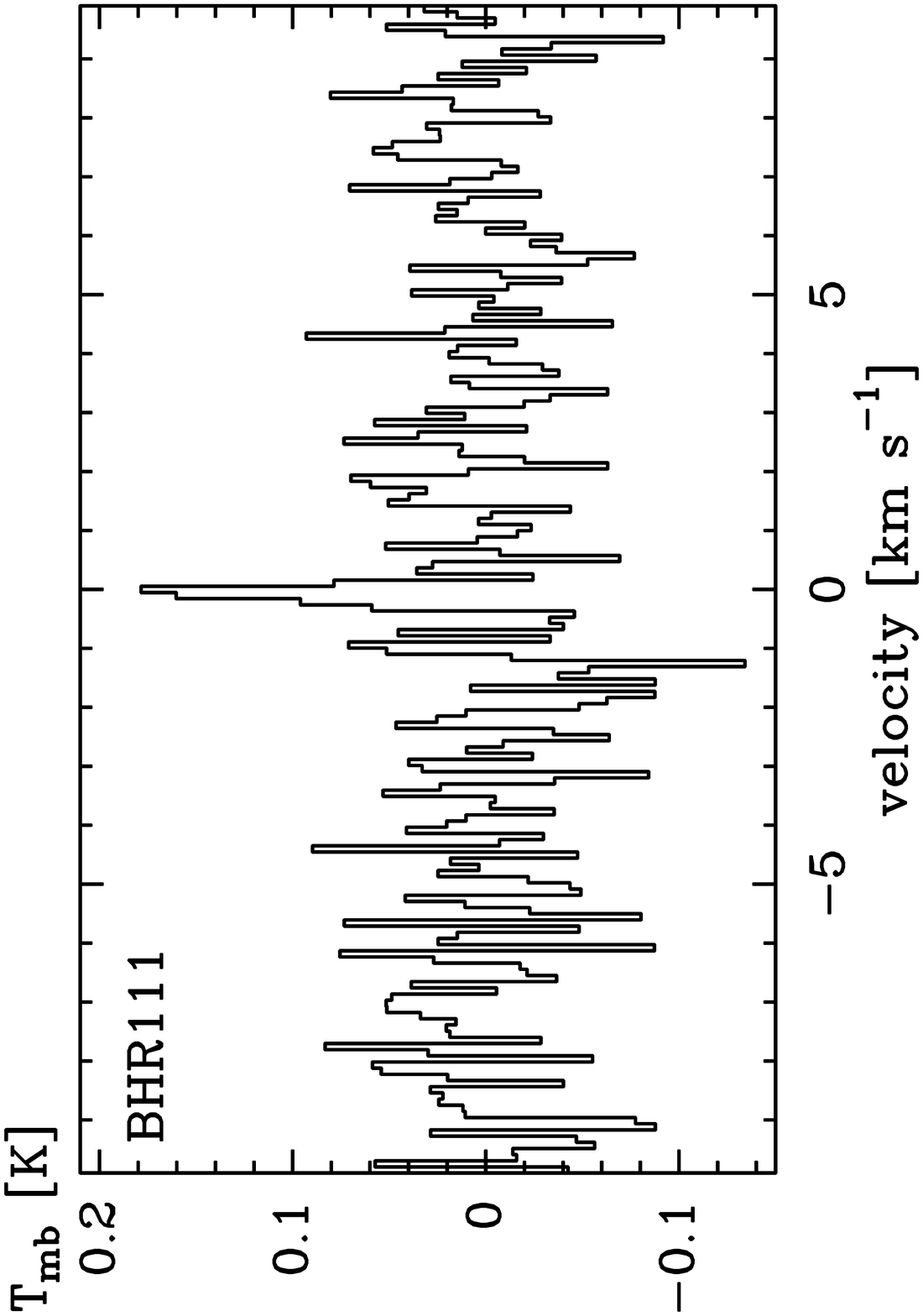}\hfill\includegraphics[angle=270]{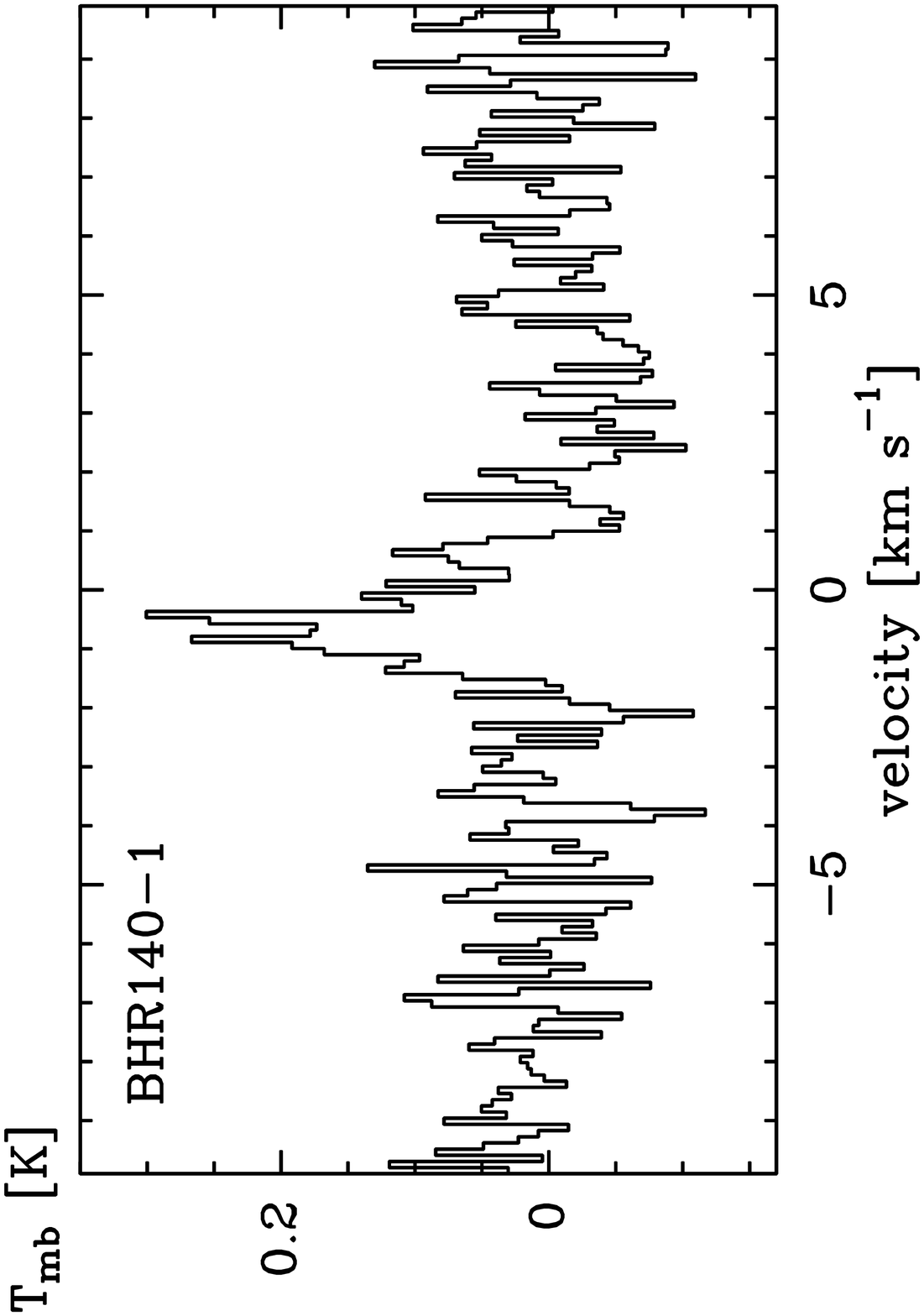}}}\hfill\parbox{0.45\hsize}{\caption{Spectra for the globules with detected CCS($2_1$--$1_0$) emission. For each source a velocity window of ca.~20~km\,s$^{-1}$ around the line is displayed. The derived line parameters are given in Table~\ref{table_results}.}
\label{spectra}}
\end{figure*}
% -----------------------------------------------------------------------------------------------

%%%%%%%%%%%%%%%%%%%%%%%%%%%%%%%%%%%%%%%%%%%%%%%%%%%%%%%%%%%%%%%%%%%%%%%%%%%%%%%%%%%%%%%%%%%%%%%%%%%%%%%%%%%%
\section{Observations}
\label{observations}
Observations in CCS($2_1$--$1_0$) of all objects from the \citet{CB88} catalog listed in Table~\ref{table_overview}, as well as NH$_3$(1,1) and (2,2) observations for three globules, have been carried out with the Effelsberg 100-m telescope of the Max-Planck-Institut f\"ur Radioastronomie (MPIfR) during March 14 and 15, 1999.
A maser receiver in the primary focus was used with a 1024 channel autocorrelator. The system temperature was in the range of 80 and 110~K. For the CCS($2_1$--$1_0$) line at 22.344~GHz the total bandwidth of 6.25~MHz of the spectrometer was used, providing a velocity resolution of 0.08~km\,s$^{-1}$. The autocorrelator was split into 2$\times$512 channels for the observations of the NH$_3$(1,1) and (2,2) lines at 23.694~GHz and 23.722~GHz, respectively, corresponding to a velocity resolution of 0.15~km\,s$^{-1}$. At the used frequencies, the beam diameter (FWHP) was 40\arcsec \,. Frequency switching was used with a total integration time per source of 30--60~min for the CCS line and 5--30~min for the NH$_3$ lines. The pointing was checked every 2 hours by cross-scans on nearby quasars. Typically, the pointing accuracy was better than 5\arcsec \ at similar elevations. The focus was checked on strong continuum sources at the beginning of the observing run and after sunset and sunrise.
The data were calibrated using cross scans on continuum sources with known flux density (Ott et al. 1994). As a primary calibrator, NGC\,7027 was used. The antenna temperatures were converted to a main beam brightness temperature $T_\mathrm{mb}$ by correcting with the elevation dependent gain, the beam and aperture efficency. The calibration is estimated to be accurate within $\sim$15\%.

The 21 sources from the catalog of \citet{Bourke95a} have been observed with the Parkes 64m telescope during March 2010 in the 
CCS($2_1$--$1_0$) transition, and ten of them additionally in NH$_3$(1,1) -- namely those sources, where the core or YSO of interest was not included in the beam of \citetalias{Bourke95b}. The 13MM receiver and a digital filterbank with a bandwidth of 8~MHz and 2048 channels were employed, resulting in a velocity resolution of 0.05~km\,s$^{-1}$. The beam diameter was 1\arcmin\, and the pointing was estimated to be accurate within 20\arcsec.
Except of BHR\,15, BHR\,23 and BHR\,28 which were observed in the position switching mode, frequency switching with a throw of 1~MHz for the CCS and 0.3~MHz for the NH$_3$ line, respectively, was used with integration times of 20--60~min. Typical system temperatures ranged from 70 to 130~K. 
After correction of the elevation-dependent gain, we applied relative calibration by means of repeated observations of BHR\,71 compared to the measurements of \citetalias{Bourke95b}. Hereby, it is ensured that the scale of our measurements and those of \citetalias{Bourke95b}, to be used for calculation of the $N_\mathrm{NH_3}$/$N_\mathrm{CCS}$ ratio, match. Although with this approach the absolute accuracy of our calibration is expected to be not optimal, comparison of the CCS observations for CB\,28, which have been carried out in Parkes and Effelsberg, show consistency.

The spectra for all globules showing (possible) detections in the CCS line are displayed in Fig.~\ref{spectra}.

%%%%%%%%%%%%%%%%%%%%%%%%%%%%%%%%%%%%%%%%%%%%%%%%%%%%%%%%%%%%%%%%%%%%%%%%%%%%%%%%%%%%%%%%%%%%%%%%%%%%%%%%%%%%

\begin{table*}
\caption{Derived CCS($2_1$--$1_0$) line parameters and column densities for CCS and ammonia, and resulting abundance ratio $N_\mathrm{NH_3}$/$N_\mathrm{CCS}$.}
\label{table_results}
\centering
\begin{tabular}{lccccccccc}
\hline \hline
Source	& Group	& $\upsilon_\mathrm{LSR}$ & $T_\mathrm{mb}$ \ \ \ & $\Delta \upsilon _\mathrm{obs}$ & $\int T_\mathrm{mb}\mathrm{d}\upsilon$ & $\Delta \upsilon_\mathrm{nth}$ & $N_\mathrm{CCS}$ & $N_\mathrm{NH_3}$ (Ref.) & $N_\mathrm{NH_3}$/$N_\mathrm{CCS}$\\
 	& & [km\,s$^{-1}$] \ \     & [mK] \ \ \    & [km\,s$^{-1}$]      & [mK~km\,s$^{-1}$] & [km\,s$^{-1}$] & [10$^{12}$\,cm$^{-2}$] & [10$^{14}$\,cm$^{-2}$] & \\
\hline
CB\,3   & 0		& \ldots & $<$\,96	& \ldots & \ldots & \ldots & $<$\,1.42 & 1.54\,$\pm$\,0.26* (1) & $>$\,104 \\
CB\,6   & I 		& \ldots & $<$\,90	& \ldots & \ldots & \ldots & $<$\,1.32 & 0.29\,$\pm$\,0.07* & $>$\,22 \\
CB\,12  & $-$I~~\, 	& \ldots & $<$\,117	& \ldots & \ldots & \ldots & $<$\,1.70 & 1.45\,$\pm$\,0.24* (1) & $>$\,86 \\
CB\,17  & \,$-$I~\tablefootmark{a} & $-$4.65\,$\pm$\,0.02 & 233\,$\pm$\,37 & 0.45\,$\pm$\,0.06 & 112\,$\pm$\,12 & 0.44\,$\pm$\,0.06 & 4.89\,$\pm$\,0.53 & 13.84\,$\pm$\,3.74 & 201\,$\pm$\,47~~  \\
CB\,22  & $-$I~~\,	& $-$1.03\,$\pm$\,0.01	& 233\,$\pm$\,37 & 0.15\,$\pm$\,0.05 & ~~36\,$\pm$\,7 & 0.08\,$\pm$\,0.08 & 1.42\,$\pm$\,0.29 & 
	3.64\,$\pm$\,0.87 (1) & 257\,$\pm$\,80\\
CB\,23  & $-$I~~\, 	& \ 4.86\,$\pm$\,0.01	& 375\,$\pm$\,39 & 0.17\,$\pm$\,0.02 & ~~66\,$\pm$\,8 & 0.11\,$\pm$\,0.03 & 2.39\,$\pm$\,0.28 & 
	10.1\,$\pm$\,1.1 (1) & 422\,$\pm$\,67\\
CB\,26  & I 	& \ldots & $<$\,114 & \ldots  & \ldots & \ldots & $<$\,1.67 & 0.48\,$\pm$\,0.18* (1) & $>$\,28 \\
CB\,28  & $-$I~~\, &  \ 9.01\,$\pm$\,0.02 & 223\,$\pm$\,40 & 0.29\,$\pm$\,0.04 & ~~69\,$\pm$\,9 & 0.26\,$\pm$\,0.05 & 3.48\,$\pm$\,0.45 & 
	1.95\,$\pm$\,1.18 (1) & 56\,$\pm$\,34 \\
CB\,34  & 0 	&  \ 0.17\,$\pm$\,0.05 & 119\,$\pm$\,37 & 0.60\,$\pm$\,0.14 & ~~57\,$\pm$\,10  & 0.59\,$\pm$\,0.15 & 2.64\,$\pm$\,0.53 & 
	0.92\,$\pm$\,0.44* (1) & 34\,$\pm$\,17 \\
CB\,44  & ($-$I)~~\, 	& \ldots & $<$\,132	& \ldots & \ldots & \ldots & $<$\,1.95 & 0.65\,$\pm$\,0.15* (1) & $>$\,34 \\
CB\,68  & 0 		& \ldots & $<$\,120 & \ldots & \ldots & \ldots & $<$\,1.48 & 12.7\,$\pm$\,0.9 (1) & $>$\,860 \\
CB\,125 & (I) 		& \ldots & $<$\,132	& \ldots  & \ldots & \ldots & $<$\,1.88 & 6.01\,$\pm$\,1.63 (1) & $>$\,320 \\
CB\,179 & $-$I~~\,	& \ldots & $<$\,102	& \ldots & \ldots & \ldots & $<$\,1.49 & 0.59\,$\pm$\,0.15* (1) & $>$\,40 \\
CB\,188 & I 	&  \ 6.14\,$\pm$\,0.04 & 111\,$\pm$\,35 & 0.44\,$\pm$\,0.08 & ~~52\,$\pm$\,10 & 0.42\,$\pm$\,0.09 & 2.21\,$\pm$\,0.41 & 1.10\,$\pm$\,0.33* (1) & 50\,$\pm$\,17 \\
CB\,222 & $-$I~~\,	& $-$2.70\,$\pm$\,0.05 & 102\,$\pm$\,25 & 0.34\,$\pm$\,0.12 & ~~36\,$\pm$\,12 & 0.31\,$\pm$\,0.14 & 1.54\,$\pm$\,0.53 & 0.65\,$\pm$\,0.08* & 42\,$\pm$\,15 \\
CB\,224 & ~~~0~\tablefootmark{b}	& \ldots & $<$\,144 & \ldots  & \ldots & \ldots & $<$\,2.61 & 6.23\,$\pm$\,1.17 (1) & $>$\,239 \\
CB\,230 & ~~~I~\tablefootmark{c}	&  \ 2.41\,$\pm$\,0.03 & 190\,$\pm$\,49 & 0.33\,$\pm$\,0.12 & ~~66\,$\pm$\,16 & 0.30\,$\pm$\,0.13 & 2.79\,$\pm$\,0.66 & 0.71\,$\pm$\,0.24* (1) & 26\,$\pm$\,9 \\
CB\,232 & \,$-$I~\tablefootmark{d}	& 12.50\,$\pm$\,0.03 & 138\,$\pm$\,41 & 0.23\,$\pm$\,0.07 & ~~34\,$\pm$\,9 & 0.19\,$\pm$\,0.09 & 1.42\,$\pm$\,0.38 & 0.54\,$\pm$\,0.21* (1) & 38\,$\pm$\,15 \\
CB\,243 & \,$-$I~\tablefootmark{e}& \ldots & $<$\,114 & \ldots & \ldots & \ldots & $<$\,1.67 & 2.77\,$\pm$\,1.04 (1) & $>$\,166 \\
CB\,244 & 0 		&  \ 4.18\,$\pm$\,0.03 & 162\,$\pm$\,41 & 0.31\,$\pm$\,0.07 & ~~54\,$\pm$\,10  & 0.29\,$\pm$\,0.07 & 1.96\,$\pm$\,0.37 &
	  8.08\,$\pm$\,2.37** (1) & 412\,$\pm$\,143\\
CB\,246 & $-$I~~\, 	& $-$0.65\,$\pm$\,0.01 & 462\,$\pm$\,39 & 0.24\,$\pm$\,0.03 & 120\,$\pm$\,11  & 0.21\,$\pm$\,0.03 & 5.67\,$\pm$\,0.49 &
	 7.70\,$\pm$\,1.01 (1) & 136\,$\pm$\,18~ \\
BHR\,12   & I+0~\tablefootmark{f}\, & \ 6.36\,$\pm$\,0.07 & 190\,$\pm$\,68 & 0.35\,$\pm$\,0.15 & ~~71\,$\pm$\,23 & 0.33\,$\pm$\,0.16 & 2.75\,$\pm$\,0.88 & 5.4 (2) & 196\,$\pm$\,62 \\
BHR\,13   & (I)		& \ldots & $<$\,294 & \ldots & \ldots & \ldots & $<$\,7.14 & 0.68\,$\pm$\,0.22* & $>$\,10 \\
BHR\,15   & ($-$I)~~\,	& \ldots & $<$\,221 & \ldots & \ldots & \ldots & $<$\,5.40 & 0.7 (2) & $>$\,13 \\
BHR\,23   & (I) 	& \ldots & $<$\,206 & \ldots & \ldots & \ldots & $<$\,5.01 & 4.07\,$\pm$\,0.67* (2) & $>$\,81 \\
BHR\,28   & ($-$I)~~\, 	& \ldots & $<$\,209 & \ldots & \ldots & \ldots & $<$\,5.08 & 3.2 (2) & $>$\,63 \\
BHR\,34   & 0 		& \ldots & $<$\,244 & \ldots & \ldots & \ldots & $<$\,5.95 & 0.91\,$\pm$\,0.50* & $>$\,15 \\
BHR\,36   & I 		& \ 5.23\,$\pm$\,0.11 & 125\,$\pm$\,61 & 0.94\,$\pm$\,0.17 & 124\,$\pm$\,26 & 0.93\,$\pm$\,0.17 & 4.13\,$\pm$\,0.86 & 7.22\,$\pm$\,2.01 &  175\,$\pm$\,60 \\
BHR\,41   & ~~~I~\tablefootmark{g} & \ldots & $<$\,198 & \ldots & \ldots & \ldots & $<$\,5.07 & 1.02\,$\pm$\,0.54* & $>$\,20 \\
BHR\,55   & ~~~0~\tablefootmark{h} & $-$5.36\,$\pm$\,0.05 & 173\,$\pm$\,93 & 0.27\,$\pm$\,0.10 & 100\,$\pm$\,62 & 0.25\,$\pm$\,0.11 & 3.38\,$\pm$\,2.11 & 5.30\,$\pm$\,1.45 & 157\,$\pm$\,107\\
BHR\,59   & (0)		& \ldots & $<$\,221 & \ldots & \ldots & \ldots & $<$\,5.37 & 2.19\,$\pm$\,0.32* (2) & $>$\,41 \\
BHR\,71   & ~~~0~\tablefootmark{i}& $-$4.52\,$\pm$\,0.08 & 209\,$\pm$\,78 & 0.87\,$\pm$\,0.14 & 193\,$\pm$\,33 & 0.86\,$\pm$\,0.14 & 6.63\,$\pm$\,1.15 & 12.25\,$\pm$\,2.75 & 185\,$\pm$\,51 \\
BHR\,74   & ($-$I)~~\,	& $-$5.89\,$\pm$\,0.09 & 111\,$\pm$\,45 & 0.52\,$\pm$\,0.22 & ~~61\,$\pm$\,18 & 0.50\,$\pm$\,0.22 & 2.19\,$\pm$\,0.66 & 1.3 (2) & 59\,$\pm$\,18\\
BHR\,86   & 0		& \ldots & $<$\,264 & \ldots & \ldots & \ldots & $<$\,7.23 & 5.0 (2) & $>$\,69 \\
BHR\,111  & (0)		& $-$0.07\,$\pm$\,0.03 & 186\,$\pm$\,44 & 0.30\,$\pm$\,0.06 & ~~59\,$\pm$\,12 & 0.28\,$\pm$\,0.07 & 2.42\,$\pm$\,0.48 & 3.6 (2) & 149\,$\pm$\,29 \\
BHR\,121-1& I		& \ldots & $<$\,212 & \ldots & \ldots & \ldots & $<$\,5.14 & 0.42\,$\pm$\,0.23* & $>$\,8 \\
BHR\,121-2& I		& \ldots & $<$\,201 & \ldots & \ldots & \ldots & $<$\,4.90 & 0.48\,$\pm$\,0.35* & $>$\,10 \\
BHR\,137  & (0)		& \ldots & $<$\,245 & \ldots & \ldots & \ldots & $<$\,5.95 & 4.3 (2) & $>$\,72 \\
BHR\,138  & 0		& \ldots & $<$\,223 & \ldots & \ldots & \ldots & $<$\,5.41 & $<$\,0.46* & \ldots \\
BHR\,139-1& I		& \ldots & $<$\,211 & \ldots & \ldots & \ldots & $<$\,5.12 & $<$\,0.45* & \ldots \\
BHR\,140-1& 0		& $-$0.63\,$\pm$\,0.07 & 221\,$\pm$\,55 & 1.21\,$\pm$\,0.16 & 285\,$\pm$\,31 & 1.21\,$\pm$\,0.16 & 11.1\,$\pm$\,1.20 & 20.4 (2) & 184\,$\pm$\,20 \\
BHR\,148  & 0		& \ldots & $<$\,282 & \ldots & \ldots & \ldots & $<$\,3.52 & $<$\,0.80* & \ldots \\
\hline
\end{tabular}
\tablefoot{(*) Optical thin approximation used to derive $N_\mathrm{NH_3}$, (**) $N_\mathrm{NH_3}$ derived from available map and map-averaged $T_\mathrm{rot}$ and $T_\mathrm{ex}$, since NH$_3$ peak lies outside the CCS beam.
\tablefoottext{a}{Main core is prestellar (SMM1+2), but there is a low-luminosity Class~I source 20\arcsec\, away, slightly outside the beam.}
\tablefoottext{b}{Main core is Class~0 (SMM1), but there is a Class~I source 30\arcsec\, away (not inside the telescope beam).}
\tablefoottext{c}{Contains two Class~I IR sources 10\arcsec\, apart.}
\tablefoottext{d}{Main core is most likely prestellar (SMM), but there is a Class~I source 12\arcsec\, away.}
\tablefoottext{e}{Main core is most likely prestellar (SMM1), but there is a Class~I source 10\arcsec\, away.}
\tablefoottext{f}{A Class~0 and Class~I YSO with cores of similar mass separated by 20\arcsec\, inside the beam.}
\tablefoottext{g}{Two close (4\arcsec\, separation) infrared sources of Class~I \citep{Santos98}.}
\tablefoottext{h}{Two close lines with comparable width $\approx$0.27~km\,s$^{-1}$ in CCS.}
\tablefoottext{i}{Double protostellar core with two IR sources 17\arcsec\, apart.}
}
\tablebib{(1)~\citet{L96}; (2)~\citet{Bourke95b}.}
\end{table*}

\begin{table*}
\caption{Results of the NH$_3$(1,1) and NH$_3$(2,2) observations.}
\label{table_nh3}
\centering
\begin{tabular}{l c c c c c c c c}
\hline \hline
Source	& $\upsilon_{\mathrm{LSR}(1,1)}$ & $T_{\mathrm{mb}(1,1)}$ & $\Delta \upsilon _{\mathrm{obs}(1,1)}$ & $\upsilon_{\mathrm{LSR}(2,2)}$ & $T_{\mathrm{mb}(2,2)}$ & $\Delta \upsilon _{\mathrm{obs}(2,2)}$ & $\tau_{\mathrm{m}(1,1)}$ & $T_\mathrm{ex}$ \\
 & [km\,s$^{-1}$] & [K] & [km\,s$^{-1}$] & [km\,s$^{-1}$] & [K] & [km\,s$^{-1}$]  & [K] & [K] \\
\hline
CB\,6      &$-$12.45\,$\pm$\,0.09 \ \ & 0.15\,$\pm$\,0.05 & 0.62\,$\pm$\,0.18 & \ldots & $<$\,0.14 & \ldots & \ldots & 6* \\
CB\,17     & $-$4.65\,$\pm$\,0.01 \  & 1.71\,$\pm$\,0.10 & 0.34\,$\pm$\,0.02 & $-$4.70\,$\pm$\,0.06 & 0.17\,$\pm$\,0.04 & 0.79\,$\pm$\,0.18 & 2.8\,$\pm$\,0.6 & 5.4\,$\pm$\,0.4 \\
CB\,222    & $-$2.71\,$\pm$\,0.04 \ & 0.29\,$\pm$\,0.05 & 0.72\,$\pm$\,0.10 & \ldots & $<$\,0.11 & \ldots & \ldots & 6* \\
BHR\,13    & 6.44\,$\pm$\,0.06 & 0.30\,$\pm$\,0.07 & 1.25\,$\pm$\,0.18 & \ldots & \ldots & \ldots & \ldots & 6*  \\
BHR\,34    & 4.78\,$\pm$\,0.04 & 0.50\,$\pm$\,0.10 & 0.69\,$\pm$\,0.11 & \ldots & \ldots & \ldots & \ldots & 6*  \\
BHR\,36    & 5.27\,$\pm$\,0.02 & 1.21\,$\pm$\,0.07 & 0.52\,$\pm$\,0.04 & \ldots & \ldots & \ldots & 1.9\,$\pm$\,0.5 & 4.3\,$\pm$\,0.2 \\
BHR\,41    & 5.46\,$\pm$\,0.05 & 0.36\,$\pm$\,0.08 & 0.92\,$\pm$\,0.16 & \ldots & \ldots & \ldots & \ldots & 6*  \\
BHR\,55    & $-$5.50\,$\pm$\,0.01 \ & 1.45\,$\pm$\,0.11 & 0.51\,$\pm$\,0.04 & \ldots & \ldots & \ldots & 1.2\,$\pm$\,0.3 & 5.1\,$\pm$\,0.4 \\
BHR\,71    & $-$4.39\,$\pm$\,0.02 \ & 3.56\,$\pm$\,0.11 & 0.71\,$\pm$\,0.03 & \ldots & \ldots & \ldots & 1.5\,$\pm$\,0.3 & 8.0\,$\pm$\,0.6 \\
BHR\,121-1 &$-$12.61\,$\pm$\,0.04 \ \ & 0.26\,$\pm$\,0.06 & 0.61\,$\pm$\,0.08 & \ldots & \ldots & \ldots & \ldots & 6* \\
BHR\,121-2 &$-$12.52\,$\pm$\,0.05 \ \ & 0.27\,$\pm$\,0.07 & 0.68\,$\pm$\,0.15 & \ldots & \ldots & \ldots & \ldots & 6* \\
BHR\,138   & \ldots & $<$\,0.21 & \ldots & \ldots & \ldots & \ldots & \ldots & 6* \\
BHR\,139-1 & \ldots & $<$\,0.20 & \ldots & \ldots & \ldots & \ldots & \ldots & 6* \\
BHR\,148   & \ldots & $<$\,0.36 & \ldots & \ldots & \ldots & \ldots & \ldots & 6* \\
\hline
\end{tabular}
\tablefoot{(*) $T_\mathrm{ex}$ is adopted as average value from \citet{L96} and \citet{Bourke95b}.}
\end{table*}

%%%%%%%%%%%%%%%%%%%%%%%%%%%%%%%%%%%%%%%%%%%%%%%%%%%%%%%%%%%%%%%%%%%%%%%%%%%%%%%%%%%%%%%%%%%%%%%%%%%%%%%%%%%%

\section{Results}
\label{results}
The results of the observations are summarized in Tables~\ref{table_results} and \ref{table_nh3}. 
Out of the 42 globules observed, thirteen were detected in the CCS($2_1$--$1_0$) line with a signal-to-noise ratio of at least three. Additional five globules show emission at a slightly lower significance level, but we believe them to be real detections since their LSR velocities are in agreement with those of ammonia lines observed by \citetalias{Bourke95b}. In contrast, NH$_3$(1,1) remained undetected only towards three globules of our sample.

From the further discussion we exclude CB\,3, CB\,34 and BHR\,23 because they likely contain clusters or high-mass star-forming regions, as well as BHR\,13, BHR\,59 and BHR\,111 because of their very uncertain evolutionary stage, and BHR\,12 because of the presence of two equally massive sources in different evolutionary groups (the source is however included in Figs.~\ref{column_density} and \ref{ratio_vs_tbol}).
Considering only the globules with fairly reliable evolutionary group (i.e. group given without brackets in Table~\ref{table_results}), CCS emission is present in 70\% (seven out of ten objects) of the globules belonging to the starless or prestellar globules of group~$-$I, in 40\% of the objects of group~0 (clouds containing Class~0 protostars; four detections) and 33\% of the group~I objects (YSOs of Class~I or later embedded; three out of nine detected). The overall detection rate (43\%) of CCS($2_1$--$1_0$) in our sample is therefore larger than found in some earlier studies of low-mass star-forming regions, globules and dark clouds (about 18--33\% in the studies of \citealt{SC96}, \citetalias{dGM06}, and \citealt{Suzuki92}), and closer to the 50\% detection rate of a survey of dense cores in the Perseus Molecular Cloud \citep{Rosolowsky2008,F09}.
For ammonia, strong emission in the sense that the hyperfine structure of the NH$_3$(1,1) transition is detected, is present in 80\% of the objects from group~$-$I, and in 70\% and 22\% of the objects of group~0 and I, respectively.

%%%%%%%%%%%%%%%%%%%%%%%%%%%%%%%%%%%%%%%%%%%%%%%%%%%%%%%%%%%%%%%%%%%%%%%%%%%%%%%%%%%%%%%%%%%%%%%%%%%%%%%%%%%%
\subsection{Analysis of CCS lines}
In Table~\ref{table_results}, Column~(1) lists the object number from the \citet{CB88} or \citet{Bourke95a} catalog, (2) the assigned evolutionary group (cf. Sect.~\ref{sample}), (3) the LSR-velocity, (4) the main beam brightness temperature, (5) the observed line width (FWHM) including instrumental broadening (of 0.08~km\,s$^{-1}$ and 0.05~km\,s$^{-1}$ for the CB and BHR sources, respectively), and in (6) the integrated intensity of the line. Column~(7) contains the nonthermal linewidths, (8) the calculated total column densities for CCS, (9) the ammonia column densities from own observations or collected from the literature, and (10) the abundance ratio $N_\mathrm{NH_3}$/$N_\mathrm{CCS}$. 
Source velocity $\upsilon_\mathrm{LSR}$, linewidth $\Delta \upsilon_\mathrm{obs}$, peak intensity $T_\mathrm{mb}$ and integrated intensity $\int T_\mathrm{mb}\mathrm{d}\upsilon$ were derived by Gaussian fits of the lines using CLASS\footnote{part of the program package GILDAS (Grenoble Image and Line Data Analysis Software), see http://www.iram.fr/IRAMFR/GILDAS/ .}. 
For the error of $T_\mathrm{mb}$, the root mean square (rms) noise of the spectrum, and for the other quantities the standard deviations of the Gaussian fits to the lines are listed in Table~\ref{table_results}. The average rms noise level of the obtained spectra is 38~mK for the CB sources and 72~mK for the BHR sources. 
As upper limits for the non-detections, a peak intensity of 3~rms is given. In the case of CB\,34 and BHR\,55 the observed CCS line was not well approximated by a Gaussian profile, therefore the value in Table~\ref{table_results} represents the intensity integrated under the actual line rather than under the fitted profile.

The total column densities $N_\mathrm{CCS}$ were calculated under the assumption of a local thermodynamic equilibrium (LTE):
\begin{equation}
N = \frac{8\pi\nu_{ij}^2}{c^2} \frac{g_0}{g_j} \frac{Q(T_\mathrm{ex})}{A_{ji}} \frac{e^{E_j/kT_\mathrm{ex}}}{\left(e^{h\nu_{ij}/kT_\mathrm{ex}} - 1 \right)} \int \tau_{\nu} \mathrm{d}\nu \quad,
\end{equation}
where $\nu_{ij}$ denotes the frequency of the transition, $g_0$ and $g_j$ are the statistical weights of the ground state and of the upper rotational level, respectively, $E_j$ is the energy of the upper level, $Q(T_\mathrm{ex})$ denotes the partition function for an excitation temperature $T_\mathrm{ex}$, $A_{ji}$ is the Einstein coefficient for spontaneous emission and $\tau_\nu$ the optical depth. We assumed the emission to be optically thin, thus the relation between optical depth and main beam brightness temperature $T_\mathrm{mb}$ from the radiative transfer equation is given by
\begin{equation}
\int \tau_\nu \mathrm{d}\nu = \frac{k}{hc}  \left( \frac{1}{e^{h\nu_{ij}/kT_\mathrm{ex}} - 1} - \frac{1}{e^{h\nu_{ij}/kT_\mathrm{bg}} - 1}\right)^{-1} \int T_\mathrm{mb} \mathrm{d}\upsilon \quad,
\end{equation}
where $T_\mathrm{bg}=2.73$~K is the cosmic background temperature. Since only one rotational transition of CCS was observed, the excitation temperature could not be determined. We adopted $T_\mathrm{ex}\approx$5~K, equal to the average value for a number of dark cloud cores as found by \citet{Suzuki92}. Restricting to the measured CCS($2_1$--$1_0$) transition with $A_\mathrm{2_1-1_0}=4.33\times10^{-7}$~s$^{-1}$ and $E_{2_1}/k=1.61$~K \citep{W97}, and $Q_\mathrm{ex}$(5~K)=23.80 \citep{LC00}, the column density ensues as follows:
\begin{equation}
N_\mathrm{CCS}~\mathrm{[cm^{-2}]} = 2.92\times10^{13}  \int T_\mathrm{mb} \mathrm{d}\upsilon~\mathrm{[K~km\,s^{-1}]} \quad.
\end{equation}
A 4~K higher excitation temperature would result in a moderate increase of about 40\% for the column density.
For the not detected lines the upper limit for $\int T_\mathrm{mb}\mathrm{d}\upsilon$ was estimated as $1.06 T_\mathrm{mb} \overline{\Delta \upsilon}$, where $\overline{\Delta \upsilon}$ was taken to be the average of the detected line widths, 0.32~km\,s$^{-1}$ for the CB sources and 0.64~km\,s$^{-1}$ for the BHR sources.

%%%%%%%%%%%%%%%%%%%%%%%%%%%%%%%%%%%%%%%%%%%%%%%%%%%%%%%%%%%%%%%%%%%%%%%%%%%%%%%%%%%%%%%%%%%%%%%%%%%%%%%%%%%%
\subsection{Ammonia lines}
For most of the sources in our sample, ammonia column densities were derived in the papers of \citetalias{L96} and \citetalias{Bourke95b}. For our own ammonia observations, column densities were derived as outlined in the following. If hyperfine structure for the NH$_3$(1,1) transition was sufficiently strong, a fit with the available procedure within the CLASS package was performed to obtain optical depth of the main line $\tau_{\mathrm{m}(1,1)}$ and intrinsic linewidth $\Delta \upsilon_\mathrm{int}$. Otherwise a Gaussian profile was fitted to the main line. The derived line parameters are summarized in Table~\ref{table_nh3}, where $\Delta \upsilon _{\mathrm{obs}(1,1)}$ refers to the intrinsic linewidth for CB\,17, BHR\,36, BHR\,55 and BHR\,71, and to the width of a single Gaussian fitted to the main component otherwise.
Further analysis was performed following standard procedures and assumptions \citep[e.g.][]{Ungerechts80,Stutzki84}.
The excitation temperature $T_\mathrm{ex}$ was calculated from
\begin{equation}
T_\mathrm{ex} \approx \frac{T_{\mathrm{mb}(1,1)}}{f_\mathrm{b}\left(1 - e^{-\tau_{\mathrm{m}(1,1)}}\right)} + T_\mathrm{bg}  \quad,
\label{Tex}
\end{equation}
with the beam filling factor $f_\mathrm{b}$ and brightness temperature of the main line $T_{\mathrm{mb}(1,1)}$.
The column density in the (1,1) levels then follows as
\begin{equation}
N_{(1,1)} = \frac{4\pi^{3/2}}{\sqrt{ln 2}} \frac{\nu_{11}^3}{c^3} \frac{\Delta \upsilon_\mathrm{int}}{A_{11}} \frac{\tau_{\mathrm{m}(1,1)}}{s_\mathrm{m}} \frac{1 + e^{-h\nu_{11}/kT_\mathrm{ex}}}{1 - e^{-h\nu_{11}/kT_\mathrm{ex}}} \quad.
\end{equation}
with the rate coefficient $A_{11}=1.67\times10^{-7}$~s$^{-1}$ %equals
and frequency $\nu_{11}$ of the NH$_3$(1,1) transition; $s_\mathrm{m}$ denotes the line strength at the main component, the value of which depends on the intrinsic linewidth of the hyperfine components blended in the main line. 
The complete ammonia column density is then estimated by
\begin{equation}
\label{NH3total}
N_\mathrm{NH_3}=N_{(1,1)} \left(1 + \frac{1}{3}\: e^{\frac{22.7~\mathrm{K}}{T_\mathrm{rot}}} + \frac{5}{3}\: e^{\frac{-41.2~\mathrm{K}}{T_\mathrm{rot}}} + \frac{14}{3}\: e^{\frac{-100.3~\mathrm{K}}{T_\mathrm{rot}}} \right).
\end{equation}
Since we did not observe the NH$_3$(2,2) transition in most cases, we could derive the rotational temperature \citep[see e.g.][]{Ungerechts80} only for CB\,17 ($T_\mathrm{rot} = 8.1\pm 0.5$~K). For the other sources we adopted a rotational temperature of 10~K for normal and 13~K for cometary globules (BHR\,12, BHR\,13, BHR\,15, BHR\,28), i.e. average values from the papers of \citetalias{L96} and \citetalias{Bourke95b}, for the calculation. 

Where the hyperfine structure was weak or only the main line of the NH$_3$(1,1) transition was detectable, so that the optical depth could not be derived, the column densities were estimated in the optically thin approximation with
\begin{equation}
N_{(1,1)} = \frac{16\pi\nu_{11}^3}{c^3} \ \frac{k}{h\nu_{11}} \ \frac{1}{f_\mathrm{b} A_{11}} \ \frac{1 + e^{h\nu_{11}/kT_\mathrm{ex}}}{1 - \frac{e^{h\nu_{11}/kT_\mathrm{ex}} - 1}{e^{h\nu_{11}/kT_\mathrm{bg}} - 1}} \int T_{\mathrm{mb}(1,1)} \mathrm{d}\upsilon ,
\label{thin}
\end{equation}
where $\int T_{\mathrm{mb}(1,1)} \mathrm{d}\upsilon$ is the intensity integrated over the main component of the NH$_3$(1,1) transition.
For the excitation temperature in equation \ref{thin}, the average value found by \citetalias{L96} and \citetalias{Bourke95b}, $T_\mathrm{ex}=6$~K has been assumed. 
The total column density can then be calculated from equation~\ref{NH3total}, adopting $T_\mathrm{rot}=10$ or 13~K as mentioned above. 
Upper limits were estimated with the same approach as described for the CCS lines, with an average linewidth of the NH$_3$(1,1) main component of 0.85~km\,s$^{-1}$.
This optical thin approximation has also been applied for sources which were detected, but for which no column density was derived in the papers of \citetalias{L96} and \citetalias{Bourke95b}. Note that for CB\,230 and CB\,232 the NH$_3$ measurement positions of \citetalias{L96} are offset by about one beam from the submillimeter cores detected by \citetalias{L2010} (identical to the CCS measurement position), and hence $N_\mathrm{NH_3}$ can be  expected to underestimate the column density towards the cores.

For the calculation of the column densities given in Table~\ref{table_results} the beam filling factors $f_\mathrm{b}$ were adopted in the following way: for sources included in \citetalias{L96} and \citetalias{Bourke95b} the source sizes derived there were used to calculate the beam filling factor to be applied to our Parkes CCS measurements, assuming that CCS and ammonia emission have comparable spatial extents \citep[which is an approximation, see e.g.][]{CS98}. For the other sources, we used the average source sizes from the papers of \citetalias{L96} and \citetalias{Bourke95b} to calculate an average beam filling factor of $f_\mathrm{b}=0.81$ for the BHR, and 0.69 for the CB objects.

%%%%%%%%%%%%%%%%%%%%%%%%%%%%%%%%%%%%%%%%%%%%%%%%%%%%%%%%%%%%%%%%%%%%%%%%%%%%%%%%%%%%%%%%%%%%%%%%%%%%%%%%%%%%
\section{Discussion}
\label{discussion}
\subsection{Column densities and fractional abundances}
\label{frac_abundance}
\begin{figure}
\centering
\resizebox{\hsize}{!}{\includegraphics[angle=270]{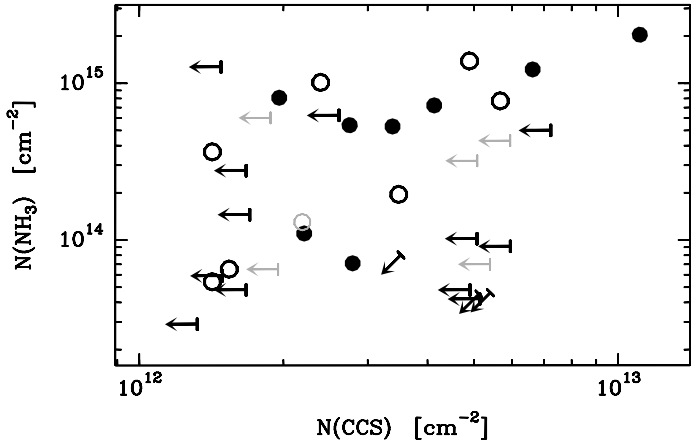}}
\caption{Column densities of NH$_3$ versus CCS. Arrows indicate upper limits; filled and open circles represent objects with evidence for ongoing star formation (group~0 and I) and no signs of current star formation (group~$-$I), respectively. Grey symbols denote sources with uncertain evolutionary group (given in brackets in Table~\ref{table_results}).}
\label{column_density}
\end{figure}

The column densities of CCS and NH$_3$, shown in Table~\ref{table_results}, do not exhibit strong dependence on the evolutionary group of the corresponding source, although the highest values for $N_\mathrm{CCS}$ are found for group~0 objects. Column densities of NH$_3$ versus those of CCS are displayed in Fig.~\ref{column_density} (without the omitted sources mentioned in Sect.~\ref{results}). The lower right part of the plot is occupied exclusively by sources undetected in CCS (i.e. only upper limits for $N_\mathrm{CCS}$); all sources with $N_\mathrm{NH_3} < 1.5\times 10^{14} \mathrm{cm}^{-2}$ have ammonia column densities obtained with the optical thin approximation. Sources with reliable detection of both molecules are found preferentially in the upper left side of the plot, as if the highest CCS column densities are present (or excited) only in globules with high ammonia column density, while lower values of $N_\mathrm{CCS}$ can be found in objects with a wider range of $N_\mathrm{NH_3}$. However, no clear distinction is visible between the distribution of the presumably younger objects of group~$-$I and the actively star-forming globules (groups 0 and I).

The large scatter of the column densities likely results from differences in the hydrogen densities of the individual clouds, therefore a comparison of fractional abundances is required for a reasonable discussion of possible evolutionary trends. 
Hydrogen column densities or masses for several globules were derived from dust emission by \citetalias{L2010}, from C$^{18}$O measurements by \citetalias{L96}, \citet{Wang95} and \citet{VB1994}, and from extinction by \citet{Racca09}. Some globules of our sample are included simultaneously in three of these studies; for these the column densities $N_\mathrm{H}$ of \citetalias{L2010} (derived by dividing the hydrogen masses from their work by the proton mass and the physical area enclosed by the 1.3\,mm dust emission in the maps) and \citetalias{L96} are found to be systematically larger than those of \citet{Wang95} by a factor of 7.6 and 2 (within 20\%), respectively. Thus, we assume the same conversion factors for the rest of the globules in the according works to bring them on the scale of \citet{Wang95}, but we do not use this approach for \citet{Racca09} and \citet{VB1994} since there is an overlap of only one source with the other studies.
For 15 globules with reliable evolutionary stage this yields fractional abundances $N_\mathrm{CCS}$/$N_\mathrm{H}$ of $7\times10^{-11}$ to  $6\times10^{-10}$ and $N_\mathrm{NH_3}$/$N_\mathrm{H}$ between $2\times 10^{-9}$ and $1\times 10^{-7}$. For CCS, the fractional abundance averaged over globules of the same evolutionary group is very similar for group~$-$I, 0 and I (about $3\times 10^{-10}$). For NH$_3$ it is highest for group~0 ($6\times 10^{-8}$) and lowest for group~I ($1\times 10^{-8}$), but this variation should not be considered significant due to the large scatter between individual globules and their small number. 
These estimated fractional abundances are comparable with those of several chemical models from the literature and an own model, which will be described later in Sect.~\ref{chem_models}, at evolutionary times close to $10^5$~yrs. Comparing with the fractional abundances observed for dense cores in cloud complexes from \citet{F09}, $N_\mathrm{NH_3}$/$N_\mathrm{H}$ from the globules is similar, but $N_\mathrm{CCS}$/$N_\mathrm{H}$ is about one order of magnitude lower than in the dense cores. However, as pointed out earlier in this section, the hydrogen column densities of the globules vary up to one order of magnitude between different studies and therefore the fractional abundances of CCS and NH$_3$ estimated here have to be considered uncertain to the same degree.
% -----------------------------------------------------------------------------------------------
% -----------------------------------------------------------------------------------------------

\subsection{NH$_3$/CCS ratio in comparison with chemical models}
\label{chem_models}
\begin{figure}
\centering
\resizebox{\hsize}{!}{\includegraphics[angle=270]{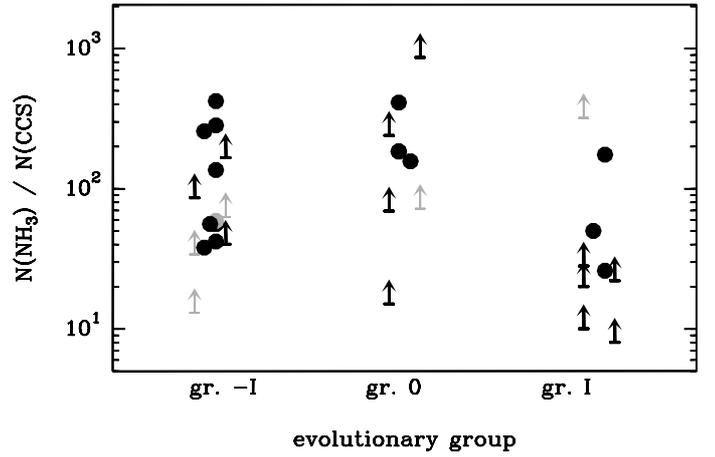}}
\caption{Abundance ratio $N_\mathrm{NH_3}$/$N_\mathrm{CCS}$ versus evolutionary group; arrows indicate lower limits and grey symbols sources with uncertain evolutionary group (slight horizontal offsets around the group positions are solely for better visibility of individual datapoints).}
\label{ratio_vs_group}
\end{figure}

In contrast, the relative abundance $N_\mathrm{NH_3}$/$N_\mathrm{CCS}$ can be discussed independent of the hydrogen densities under the assumption that both molecules trace similar cloud regions, which is likely because of comparable excitation conditions and a similar telescope beam for both observations.
Figure~\ref{ratio_vs_group} shows $N_\mathrm{NH_3}$/$N_\mathrm{CCS}$ versus evolutionary group. 
Altogether, we derive abundance ratios from about 20 to 860, while the observable range is limited to ca. $N_\mathrm{NH_3}$/$N_\mathrm{CCS}\leq 2000$ by our CCS detection limit. 

Compared to samples of earlier papers, the $N_\mathrm{NH_3}$/$N_\mathrm{CCS}$ ratios derived for isolated Bok globules are on average similar to those of dense cores in the Perseus molecular cloud \citep{Rosolowsky2008}, but a factor of two higher than those in the dark clouds studied by \citet{Suzuki92}. This corresponds to the fact that despite a comparable range of ammonia column densities in the \citet{Suzuki92} sample and our globules, we do not find extremely high CCS column densities typical of carbon-chain producing regions \citep[e.g. ][where carbon-chain producing regions are defined as having $N_\mathrm{NH_3}$/$N_\mathrm{CCS}\leq 10$]{Hirota2009}.

The number of sources is too small to allow for a detailed statistical analysis, but taking into account only sources with reliable evolutionary stage, the  following picture arises: within each evolutionary group, a relatively large range of abundance ratios $N_\mathrm{NH_3}$/$N_\mathrm{CCS}$ spanning about one order of magnitude is observed, but there is comparatively little variation of the values between the different groups. Fig.~\ref{ratio_vs_group} possibly suggests a slightly decreasing tendency of $N_\mathrm{NH_3}$/$N_\mathrm{CCS}$ going from group~0 to group~I globules, but this trend is only marginal. Thus, it can be concluded that the ratio $N_\mathrm{NH_3}$/$N_\mathrm{CCS}$ is rather similar across the Bok globules observed in this study, despite them harbouring YSOs in different evolutionary stages. 

Specifically, the $N_\mathrm{NH_3}$/$N_\mathrm{CCS}$ ratio does not increase going from the presumably youngest objects of group $-$I to the most evolved sources of group I.
This finding does not easily fit into the general picture of most chemical models, according to which CCS as an early-phase molecule is expected to decrease in abundance rapidly around $10^6$~yr, while the slowly forming ammonia reaches its maximum abundance in a later stage of the chemical evolution~-- leading to the anticipation of steadily increasing $N_\mathrm{NH_3}$/$N_\mathrm{CCS}$ ratio along with the evolutionary stage of the globules.
% -----------------------------------------------------------------------------------------------
\begin{figure}
\centering
\resizebox{\hsize}{!}{\includegraphics[angle=270]{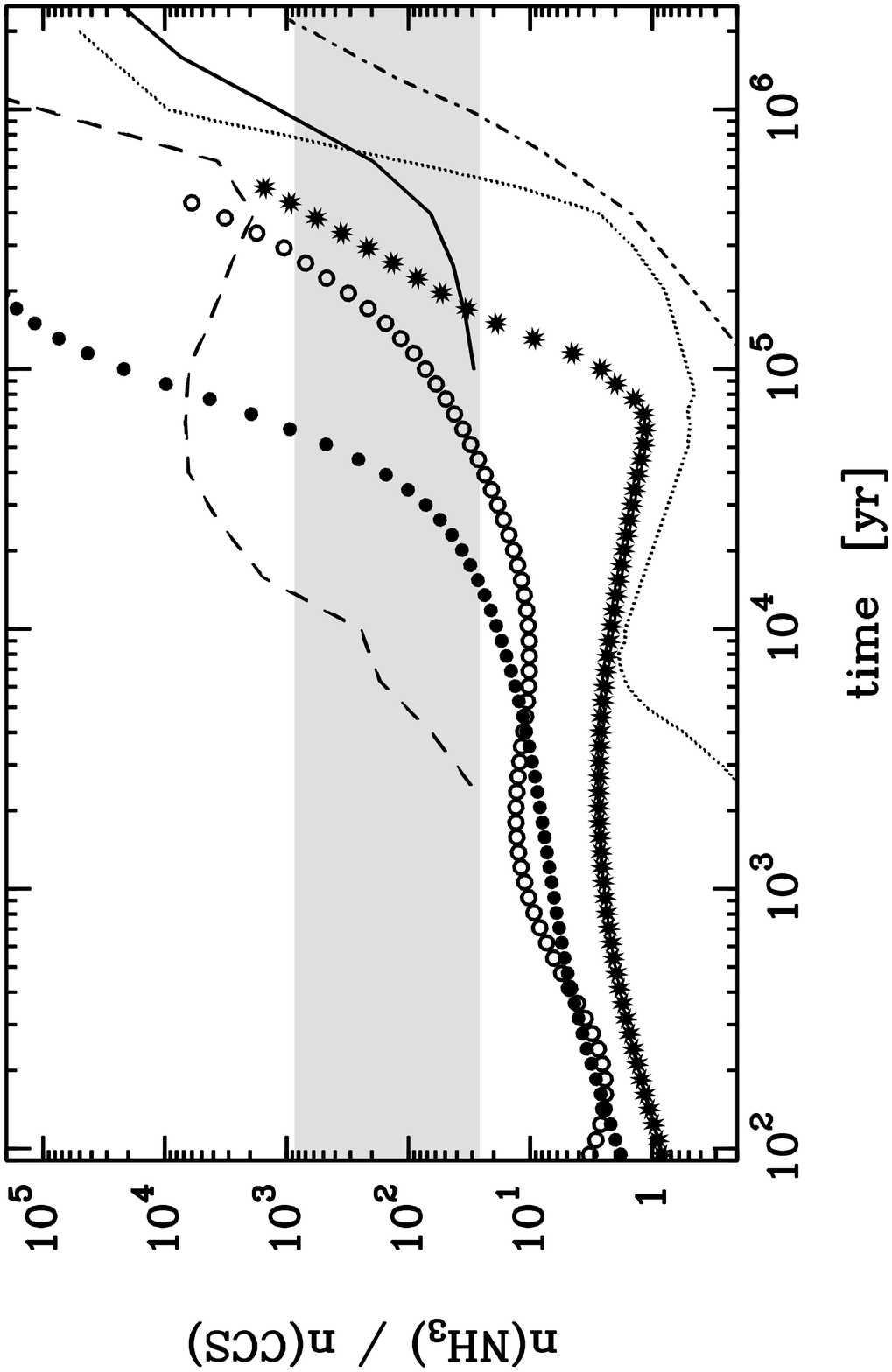}}
\caption{$\mathrm{NH_3}$/$\mathrm{CCS}$ ratio from chemical model calculations (see Sect.~\ref{chem_models}) of \citet[][dash-dotted line]{Suzuki92}, \citet[][dashed line]{Scappini1998}, \citet[][solid line]{Bergin2000} and \citet[][dotted line]{Aikawa2001}, and this work (10~K warm group~$-$I model marked with open circles, 15~K group~0 model with filled circles and 25~K group~I model with asterisks. The observed range of $N_\mathrm{NH_3}$/$N_\mathrm{CCS}$ is designated by the shaded area.}
\label{models}
\end{figure}
% -----------------------------------------------------------------------------------------------
For comparison, we show in Fig.~\ref{models} the $N_\mathrm{NH_3}$/$N_\mathrm{CCS}$ ratio with respect to the chemical age of a cloud, calculated from the evolution of CCS and NH$_3$ abundances from four chemical models in the literature: \citet[][dash-dotted line]{Suzuki92} and \citet[][dashed line]{Scappini1998} use pseudo-time-dependent chemical models in which the gas density is constant within time. 
In contrast, \citet[][solid line]{Bergin2000} and \citet[][dotted line]{Aikawa2001} take into account the dynamics of a collapsing core, as well as depletion of species from the gas phase onto, and desorption from, dust grains. However, both models do not include reactions on grain surfaces except of $\mathrm{H_2}$ formation and ion-electron recombination in \citet{Aikawa2001}.
As initial conditions all models take hydrogen in molecular form and carbon and sulphur as ions, the initial abundances are taken as those typical of diffuse clouds or depleted by a certain factor from solar abundances; \citet{Bergin2000} allow the cloud to evolve for $1.5\times 10^5$~yr at constant density and take the chemical composition after this time as initial values for the dynamically evolving core. 
All models 
consider the chemical evolution in a region at a constant temperature of 10~K and shielded from external UV radiation. %
\citet{Aikawa2001} follow the chemical evolution of an infalling fluid element in a collapse according to the Larson-Penston solution (we show here their result for a collapse slowed down by a factor of 10).  From \citet{Bergin2000} we show the model for a collapse with ambipolar diffusion and dust grains covered with a CO mantle. Both papers examine several variations of their models (e.g. grain properties, collapse timescales), which result in somewhat different time evolutions of the molecular abundances, but do not deviate significantly from the examples shown in Fig.~\ref{models}.

Despite the different approaches, all models agree upon a rapid increase of the $N_\mathrm{NH_3}$/$N_\mathrm{CCS}$ ratio by three orders of magnitude at an evolutionary time of several 10$^5$~yr, caused by a fast decrease of the CCS abundance due to depletion onto grains, destruction by reactions and missing replenishment. The first broad peak seen in the results of \citet{Scappini1998} and \citet{Aikawa2001} is due to a very slow formation and increase of the NH$_3$ abundance, while CCS is still being formed efficiently.

In general, a comparison of the absolute values observed with those from modelling has to be considered with caution. The initial conditions assumed and the exact starting point for reactions 
defined in chemical models might not necessarily be in good agreement with real globules, and certain scatter has to be expected due to the natural fluctuations of initial conditions in the variety of globules.
Nevertheless, comparing the measured values for $N_\mathrm{NH_3}$/$N_\mathrm{CCS}$ abundance ratio with the models suggests that the observed objects might all be in an evolved state matching the region after $10^5$~yr despite their different evolutionary stages. 
% -----------------------------------------------------------------------------------------------
\begin{figure}
\centering
\resizebox{\hsize}{!}{\includegraphics[angle=270]{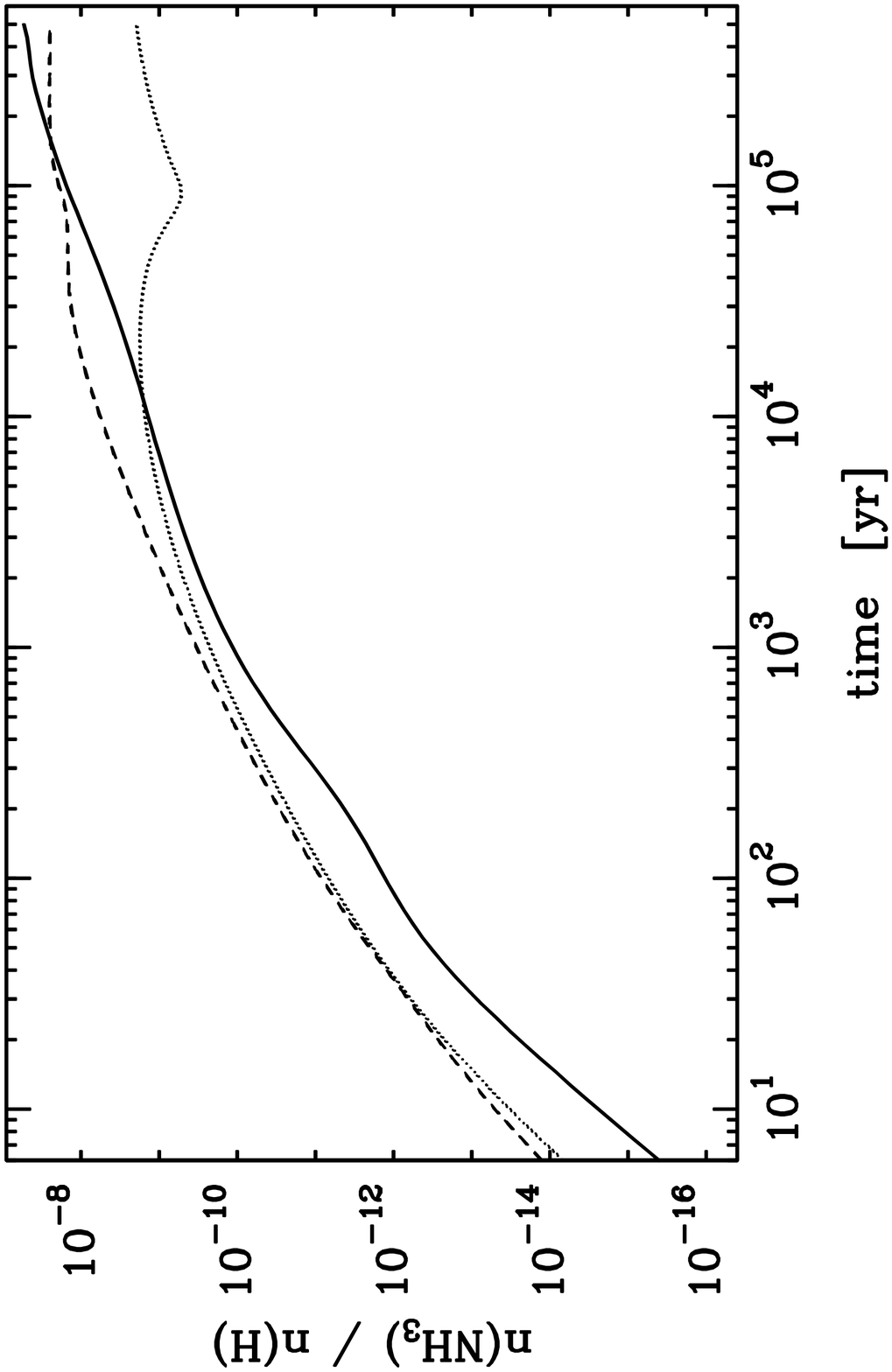}}
~~\\
\centering
\resizebox{\hsize}{!}{\includegraphics[angle=270]{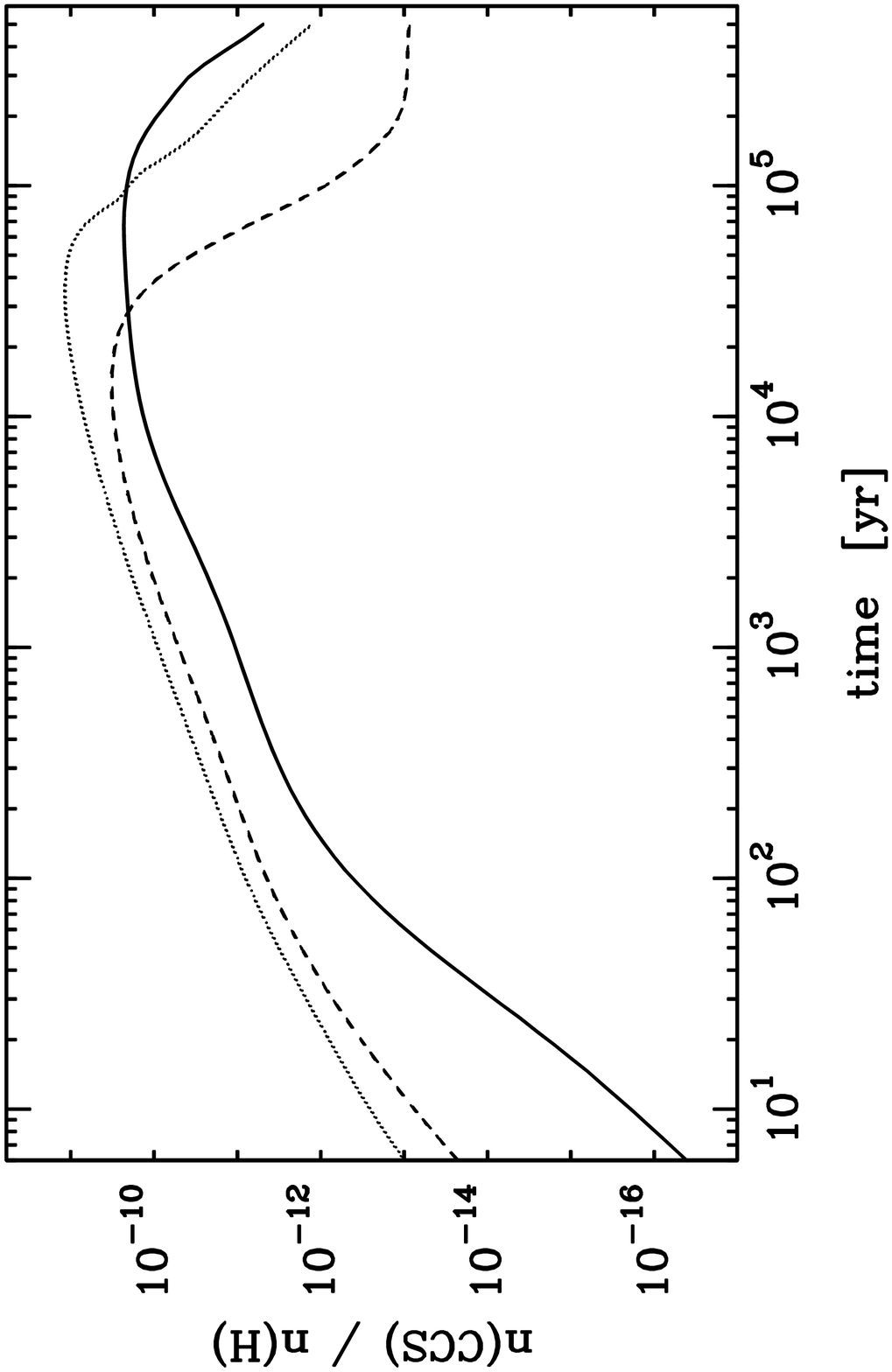}}
\caption{Upper part: Evolution of the NH$_3$ fractional abundance resulting from our model described in Sect.~\ref{chem_models} for group~$-$I model (10~K, solid line), group~0 model (15~K, dashed line) and group~I model (25~K, dotted line). Lower part: The same for CCS.}
\label{model_dima}
\end{figure}
% -----------------------------------------------------------------------------------------------
\medskip
In addition, we employed an own chemical model to calculate abundances of NH$_3$ and CCS for sources of the three groups $-$I, 0 and I. For this purpose a recent chemical network including a variety of grain surface reactions \citep{Semenov2010} was used together with a simplified physical model. Representative physical parameters for each group were taken from the globule study by \citetalias{L2010}, i.e. the density is described by a power law function of radius (with exponent between $-$1 and $-$2) in the outer region and flattening towards the core center. We used a temperature of 10~K for group $-$I, 15~K for group~0, and 25~K for group~I sources, respectively, based on observations and models of the mass-averaged dust temperature (\citetalias{L2010} and references therein).
Starting from "low metal"' atomic abundances \citep{Lee1998} as initial values, the evolution of NH$_3$ and CCS abundance relative to H was calculated separately for the typical physical conditions of each group. We refrained from constructing a piecewise model with sharp transitions from one group to the next at defined timesteps, since in reality a smooth transition of physical conditions can be expected, and the lifetimes of the different groups are not tightly constrained. Instead, we discuss qualitative differences of the three models.

The resulting evolution of fractional NH$_3$ and CCS abundances is displayed in Fig.~\ref{model_dima}.
In our chemical model, CCS forms fast in a neutral-neutral reaction of CCH and S, and reaches its maximum concentration early. NH$_3$ is formed partly on dust grains through surface hydrogenation of nitrogen, and partly in the gas phase mainly via a sequence of ion-molecule reactions. Desorption from grain surfaces occurs mainly through cosmic-ray heating and cosmic-ray-induced far-UV photons in the cold and shielded environments considered here. 
After a few $10^5$~yrs CCS is efficiently removed from the gas phase by freeze out onto dust grains or destruction by reactions with oxygen, and rapidly drops in abundance. For the 15 and 25~K warm group~0 and group~I models, prior to this late-time depletion of CCS a number of competitive surface reactions become active, which enhance the CCS (re)production and lower the NH$_3$ abundance, resulting in a temporary drop of the NH$_3$ abundance at about $10^5$~yrs and a slightly increased CCS peak abundance compared to the 10~K model of group~$-$I sources.
For other carbon-chain molecules, a similar enhancement in moderately warm environments ('warm carbon-chain chemistry') has been pointed out recently \citep[e.g.][]{Sakai08,Aikawa2008,Hassel2008}. 
These differences in chemical evolution of our models originate primarily from the different temperatures, while the difference in density profiles has smaller influence. For simplicity, we will refer to the models of the three groups ($-$I, 0 and I) by their temperature (10, 15 and 25~K) in the following. 

As a result, in the early phase the ratio $N_\mathrm{NH_3}$/$N_\mathrm{CCS}$ evolves very slowly from values close to unity to few dozen for all three models. For the 10~K model, the abundance ratio starts to increase faster around $10^5$~yrs and crosses the range observed in the Bok globules of our sample within $2.5\times10^5$~yrs. For the somewhat warmer 15~K model, $N_\mathrm{NH_3}$/$N_\mathrm{CCS}$ increases earlier and more rapidly, while for the 25~K model the abundance ratio remains low ($<10$) until $>1\times10^5$~yrs before increasing rapidly to $>10^3$ like the other two models.
In Fig.~\ref{models}, the evolution of $N_\mathrm{NH_3}$/$N_\mathrm{CCS}$ for the three parameter sets of our chemical model is designated by open and filled circles (10~K and 15~K, respectively) and asterisks (25~K).

Comparing with the observational data shown in Fig.~\ref{ratio_vs_group}, the observed range of $N_\mathrm{NH_3}$/$N_\mathrm{CCS}$ ratios could be compatible with the rapidly increasing parts of our three model curves at evolutionary times of few $10^4$--$10^5$~yrs. However, the 10~K and 15~K model are close to the lower $N_\mathrm{NH_3}$/$N_\mathrm{CCS}$ limit derived from our observations for most of their evolutionary time, while the 25~K model remains at significantly lower values up to $10^5$~yrs. Thus, if one assumes that the actual chemical ages of Bok globules may possess a certain scatter, one could also expect to find globules with very low $N_\mathrm{NH_3}$/$N_\mathrm{CCS}$ ratios ($\leq 5$) if the group~I model is applicable~-- but such low ratios are not confirmed in our sample. The range of fractional abundances estimated from the observations in Sect.~\ref{frac_abundance} agrees for NH$_3$ with both 10 and 15~K model for times $>10^4$~yrs, while the NH$_3$ abundance in the 25~K model remains slightly below the observed range for the whole time span covered by the model. For CCS, the peak abundances of 10 and 15~K model around $10^4$--$10^5$~yrs fall into the observed range, while the 25~K model exceeds it. Although having to keep in mind the uncertainties of the estimated fractional abundances, this may be another indication that the 25~K model is not in agreement with the observations, while 10 and 15~K warm model fit them equally well. 

Originally, the choice of 10--25~K for the models of the different groups was motivated by models \citep[e.g.,][]{Shirley2002,Galli2002,Bergin2006} and observations \citep[e.g. for CB\,244 in][]{Stutz2010} of the dust temperature. In general, dust temperatures are expected to be elevated ($\sim$15~K) close to the cloud surface heated by the interstellar radiation field, and lower ($\sim$10~K) in the moderately dense envelope where CO line radiation cools the gas efficiently. In the center of prestellar cores they may be as low as 5~K (depending on the external radiation field), or accordingly warmer in the case of an internal heating source (20--30~K in \citealt{Shirley2002}). 
In the dense (n$_\mathrm{H}\geq 10^5$~cm$^{-3}$) inner regions gas and dust are expected to be well coupled through collisions and hence similar in temperature. 

In contrast, the beam- (\citetalias{Bourke95b}) or map-averaged (\citetalias{L96}) NH$_3$ rotational temperatures (which represent a good estimate of the kinetic temperature at the low temperatures prevalent in globules, see e.g. \citealt{Stutzki84}) are 9--16~K for globules of all evolutionary stages (\citetalias{L96,Bourke95b}). 
This most likely results from the fact that the warm cores of Class~0 and more evolved objects have typical sizes of few thousand AU (\citetalias{L2010}), and hence comprise only a small fraction of the NH$_3$ beam area for the average globule distance of our sample. In addition, the innermost warmest and densest parts of the core may not be traced well by NH$_3$ due to high optical depth. 
Thus, with the large beam of the NH$_3$ observations considered here, cool ($\sim$10--15~K) gas from the moderately dense envelope can be expected to dominate the signal (see also \citealt{Crapsi2007}, where the central gas temperature drop in a prestellar core is detectable with interferometric, but not single-dish observations).

Moreover, a scatter of temperatures in the range 10--15~K for the bulk of gas may also result from differences in size, density and local UV background radiation for the individual globules, and not only from their evolutionary stage. 
In \citet{F09} the kinetic temperatures from NH$_3$ of protostellar and starless cores cover a similar range as those for our globules, and the slight differences between the star-forming and starless group detected by them may be discernable only for sample sizes significantly larger than ours.

If indeed, as suggested by our chemical model, small temperature differences of $\sim5$~K could already perceptibly affect the progression of the $N_\mathrm{NH_3}$/$N_\mathrm{CCS}$ ratio around $10^5$~yrs, then the superposition of $N_\mathrm{NH_3}$/$N_\mathrm{CCS}$ evolution curves for several initial temperatures could lead to a relatively large scatter of $N_\mathrm{NH_3}$/$N_\mathrm{CCS}$ ratios within one evolutionary group, while  simultaneously smearing out differences between the evolutionary groups (differences of other initial physical parameters as density may also contribute, but we do not address them here in detail since the observational data is too sparse for comparison). This would agree with the observation that the (beam-averaged) $N_\mathrm{NH_3}$/$N_\mathrm{CCS}$ ratio is rather similar across globules of all evolutionary stages and exhibits a relatively large scatter within each group. Additional contributions to the scatter of the $N_\mathrm{NH_3}$/$N_\mathrm{CCS}$ abundance ratio among globules of a single group may arise from the significant fraction of globules containing adjacent objects of different evolutionary stages (Sect.~\ref{sample}), indicative of sequential star formation, and a range of ages even among the objects compiled in one evolutionary group.

To test this hypothesis in detail, it would be necessary to disentangle the influence of gas temperature and age (i.e. evolutionary group), for which our dataset is significantly too small.
It could also be of interest to evaluate a possible increase of gas temperature and change of chemistry on small scales close to forming protostars, for which observations at much higher resolution are needed.

\subsection{Abundance ratio and bolometric temperature}
\label{Tbol}

% -----------------------------------------------------------------------------------------------
\begin{figure}
\centering
\resizebox{\hsize}{!}{\includegraphics[angle=270]{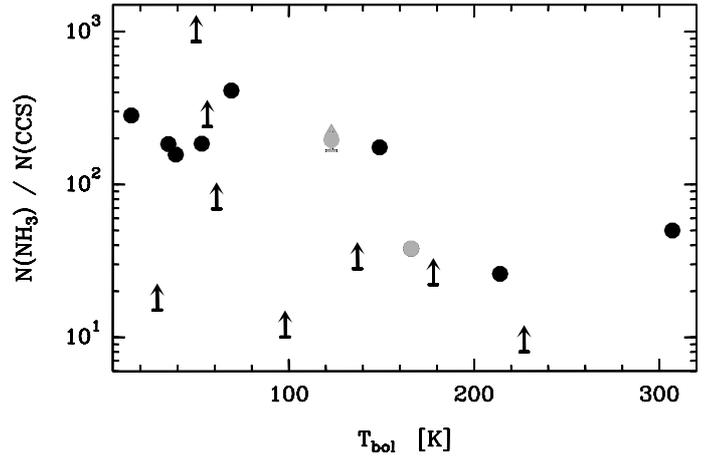}}
\caption{Abundance ratio $N_\mathrm{NH_3}$/$N_\mathrm{CCS}$ versus bolometric temperature of embedded sources. Arrows indicate lower limits of the abundance ratio, grey symbols mark $T_\mathrm{bol}$ of combined spectral energy distributions (cf. Sect.~\ref{Tbol}).}
\label{ratio_vs_tbol}
\end{figure}
% -----------------------------------------------------------------------------------------------
In the previous years it has been pointed out that the evolution of young stellar objects is well traced by the bolometric temperature $T_\mathrm{bol}$, defined as the temperature of a blackbody with the same mean frequency as the spectrum of the observed source \citep{Ladd91,Myers93}. \citet{Chen95} demonstrated a tight correlation between $T_\mathrm{bol}$ and the age of protostars and pre-main-sequence stars in particular for embedded sources. For a number of young stellar objects in globules of our sample, the bolometric temperature has been evaluated by \citetalias{L2010} and \citet{Racca09}. However, $T_\mathrm{bol}$ is mainly useful to track the evolution of Class~0 and later sources, since for most of the coldest, presumably youngest sources (e.g. CB\,246, BHR\,137) $T_\mathrm{bol}$ is not known due to a spectral energy distribution not well constrained at shorter wavelengths.

We show our measured $N_\mathrm{NH_3}$/$N_\mathrm{CCS}$ abundance ratio versus $T_\mathrm{bol}$ adopted from the works by \citetalias{L2010} and \citet{Racca09}, in Fig.~\ref{ratio_vs_tbol}. In the case of CB\,232 and CB\,243 (as well as BHR\,12 denoted in grey colour), where both a prestellar core and a Class~I source are included in the beam, $T_\mathrm{bol}$ represents the value for the combined spectral energy distribution of the adjacent sources, which is in both cases dominated by the Class~I source while the molecular emission is expected to emerge mainly from the prestellar cores. However, in both globules the separation of prestellar core and Class~I source equals the typical size of a prestellar core (8000~AU in \citetalias{L2010}); in this case the presence of an evolved YSO might influence the chemistry of the prestellar core and accordingly both may be not completely unrelated. 
For a definite judgement about this question, a mapping of the line emission at small scales would be required.

From most chemical models in the literature as shown in Fig.~\ref{models}, an increase of the $N_\mathrm{NH_3}$/$N_\mathrm{CCS}$ ratio towards warmer bolometric temperatures, which in turn are thought to represent increasingly older and more evolved sources, would be expected.
Despite the scarceness of our dataset, there are no indications of such an increase. Instead, the $N_\mathrm{NH_3}$/$N_\mathrm{CCS}$ abundance ratio shows a tentatively trend of decrease towards increasing bolometric temperature. However, due to the small number of globules with both known bolometric temperature and detected CCS emission, an interpretation is even more difficult than in Sect.~\ref{chem_models}.

\subsection{Nonthermal linewidths}
\label{linewidths}
The observed linewidths $\Delta \upsilon_\mathrm{obs}$ 
include an instrumental broadening of $\Delta\upsilon_\mathrm{instr}=0.08$~km\,s$^{-1}$ (Effelsberg) and 0.05~km\,s$^{-1}$ (Parkes) which can be subtracted under the assumption of a
Gaussian profile for the line as well as for the spectrometer:
\begin{equation}
\Delta \upsilon^2 = \Delta \upsilon_\mathrm{obs}^2 - \Delta \upsilon_\mathrm{instr}^2 \quad.
\end{equation}
In the same way, the thermal contribution to the linewidth,
\begin{equation}
\Delta \upsilon_\mathrm{therm} = \sqrt{\frac{8\,\mathrm{ln}2\,k\,T_\mathrm{kin}}{m}} \quad,
\end{equation}
with $m$ denoting the molecular mass of the molecule, was subtracted to obtain the nonthermal linewidths $\Delta\upsilon_\mathrm{nth}$ in
Table~\ref{table_results}. We used the rotational temperature $T_\mathrm{rot}$ derived by \citetalias{L96} and \citetalias{Bourke95b} from their ammonia observations as an estimate for
the kinetic temperature $T_\mathrm{kin}$. Where such values for the individual globules were not available, an average temperature of 
$T_\mathrm{kin}=10$~K was assumed. The typical thermal contribution for the CCS and NH$_3$ linewidths is 0.09 and 0.16~km\,s$^{-1}$, 
respectively.
It should be pointed out that, because the main component of the NH$_3$(1,1) line consists of several usually blended hyperfine components distributed over 0.7~km\,s$^{-1}$, linewidths derived from a single Gaussian fit to the main component are dominated by the separation of the blended hyperfine components rather than the intrinsic linewidth for small linewidths. An intrinsic linewidth can be in principle estimated from such fits, but suffers from large uncertainties for the typical small linewidths observed in Bok globules. Therefore we discuss nonthermal linewidths of only those globules with sufficiently well detected main and satellite lines of NH$_3$(1,1), for which intrinsic linewidths can be derived directly from fits of the hyperfine structure.

For both NH$_3$ and CCS, the majority of nonthermal linewidths is found in the range 0.2--0.5~km\,s$^{-1}$ for all evolutionary groups.
In addition, large nonthermal linewidths 0.6--1.2~km\,s$^{-1}$ are observed in globules of group~0 and I, while very small linewidths $\leq$0.1~km\,s$^{-1}$ are present exclusively in globules of group~$-$I. 
The nonthermal linewidth can be understood as a measure for turbulence of the medium and the existence of unresolved velocity components within the  area covered by the telescope beam. An increased linewidth $\Delta \upsilon_\mathrm{nth}$ for actively star-forming regions could be expected due to turbulence and outflow motions induced by the presence of a protostar. 
However, for globules at larger distances a physical region of proportionally larger dimension is enclosed by the beam, and could therefore result in larger linewidths. We are fairly confident that our results are not biased by this effect, since no clear relationship between linewidth and distance is discernable in our sample.

Among the starless or prestellar sources of group~$-$I, the globules with extremely small nonthermal linewidths (CB\,22 and CB\,23) are distinguished from the rest by an absence of IRAS and millimeter sources, i.e. there is no evidence for any embedded sources. In contrast, three of the group~$-$I globules with intermediate $\Delta\upsilon_\mathrm{nth}$ harbour more evolved Class~I sources in the vicinity of the prestellar core, while the remaining globules are poorly studied.
The group~0 and I sources with large nonthermal linewidths exhibit not very Gaussian-shaped profiles (BHR\,140-1, BHR\,36) or hints of at least two velocity components (BHR\,71, also BHR\,55). However, they do not differ distinctly from the rest of the group~0 and I globules in properties like presence of outflows, multiplicity, etc.

In the majority of globules, the nonthermal CCS linewidths are smaller than those of NH$_3$, but also the opposite case or similar linewidths are observed. A possible explanation could be that both molecules trace not exactly similar cloud regions, as in the cases demonstrated by \citet{SC98} and \citet{LC00}, where the most dense parts of cores were preferentially detected in NH$_3$, while CCS emission arises from a more extended surrounding region. Such central depletion holes in evolved sources have also been observed for other carbon-bearing species, e.g. for the CCH radical \citep[e.g.,][]{Beuther2008,Padovani2009}, which resembles in many properties, and is thought to be a precursor of, CCS.
Since both NH$_3$ and CCS possess a roughly comparable critical density ($\sim 10^3$--$10^4$~cm$^{-3}$), a difference in the emitting region is most likely a result of different spatial distributions. 

This in turn might mean that both molecules trace regions with somewhat different physical conditions, and also result in an incorrect estimate of the thermal component for the linewidths when assuming the same kinetic temperature for both CCS and NH$_3$.
Altogether, it is difficult to draw firm conclusions regarding the origin of the spread of nonthermal linewidths due to the sparse dataset and availability of only single-point measurements.

%%%%%%%%%%%%%%%%%%%%%%%%%%%%%%%%%%%%%%%%%%%%%%%%%%%%%%%%%%%%%%%%%%%%%%%%%%%%%%%%%%%%%%%%%%%%%%%%%%%%%%%%%%%%
\section{Summary}
\label{summary}
Observations of 42 Bok globules in CCS and partially in NH$_3$, supplemented with NH$_3$ measurements from the literature (\citetalias{L96}, \citetalias{Bourke95b}) enabled us to assess the abundance ratio of both molecules for {18} objects (and derive lower limits for additional 21 globules) in different evolutionary stages ranging from starless and prestellar globules (designated as group~$-$I) over clouds containing Class~0 YSOs (group~0) to globules harbouring young stellar objects of Class~I and later (group~I). In the following we summarize our main results:

\begin{itemize}
\item[1.] The CCS(2$_1$--$1_0$) line is detected in 18 of 42 Bok globules. The detection rate is highest in the objects of group~$-$I (70\%) and decreases towards the globules of group~0 and I (40 and 33\%, respectively). Ammonia is detected (together with literature data) in 39 globules.
\smallskip
\item[2.] We derive $N_\mathrm{NH_3}$/$N_\mathrm{CCS}$ ratios in the range 26--422, plus one lower limit of 860, for isolated Bok globules. In particular, within the limits of our survey, we find neither extremely low abundance ratios ($\leq 10$) like those typical of carbon-chain producing regions in dark cloud \citep{Suzuki92,Hirota2009}, nor extremely high abundance ratios (several $10^3$) as expected for evolved sources from chemical models for cloud cores by different authors (Fig.~\ref{models}). 
\smallskip
\item[3.] We do not observe an increase of the $N_\mathrm{NH_3}$/$N_\mathrm{CCS}$ abundance ratio going from the starless and prestellar globules towards evolved globules containing Class~I or later sources, nor from lower towards higher bolometric temperatures of the embedded YSOs, as would be expected from various chemical models from the literature and this work (Sect.~\ref{chem_models}). 
Instead, the ratio exhibits a considerable scatter but is roughly constant across all evolutionary groups (Fig.~\ref{ratio_vs_group}), with a tentatively - however, statistically not significant - decreasing trend from globules containing Class~0 protostars (group~0) towards globules with Class~I or later YSOs (group~I). 
\smallskip
\item[4.] An own chemical model (Sect.~\ref{chem_models}) indicates that a slight temperature increase (15 instead of 10~K) could affect the $N_\mathrm{NH_3}$/$N_\mathrm{CCS}$ ratio perceptibly at evolutionary times around $10^5$~yrs. Since NH$_3$ rotational temperatures vary in the same range, this may suggest that the observed roughly constant but scattered distribution of (beam-averaged) $N_\mathrm{NH_3}$/$N_\mathrm{CCS}$ ratios could result from a superposition of evolutionary tracks for different initial globule temperatures. 
In contrast, a 25~K warm model seems less likely, since neither such high gas temperatures, NH$_3$ and CCS fractional abundances in agreement with the model, nor very low $N_\mathrm{NH_3}$/$N_\mathrm{CCS}$ ratios ($\leq5$), as expected from the model even at late evolutionary times, are observed.
\smallskip
\item[5.] The smallest nonthermal linewidths of CCS ($\sim 0.1$~km\,s$^{-1}$), indicating a very low level of turbulence, are detected only in two globules of group~$-$I without any associated infrared or millimeter sources. In contrast, broad CCS lines with evidence for multiple velocity components are found only among star-forming globules (groups~0 and I). Beyond that, no firm conclusions about the nonthermal linewidths can be drawn due to the small sample size.
\end{itemize}

We conclude that our observed $N_\mathrm{NH_3}$/$N_\mathrm{CCS}$ abundance ratios derived from single-dish observations with relatively large beam, although related to the evolutionary state of the embedded objects, cannot be alone and straightforward interpreted as an evolutionary tracer for isolated Bok globules.
Another major problem hampering this study is the immediate vicinity of objects in different evolutionary stages encountered in many of the globules \citepalias[see also ][]{L2010}, which makes it difficult to assess the amount of emission arising from each source or a possible mutual influence on physical and chemical conditions.
In addition, the assumption of NH$_3$ and CCS tracing the same spatial regions is an approximation and may not hold for all globules. The distribution of CCS -- central depletion hole or enhanced in the warm core region of evolved sources -- is of particular interest.
Well-resolved mapping of the line emission of both molecules for a large sample of globules, {assessment of the gas temperature,} and a position-dependent evaluation of the $N_\mathrm{NH_3}$/$N_\mathrm{CCS}$ ratio, will be required to address the mentioned aspects in full detail.

%%%%%%%%%%%%%%%%%%%%%%%%%%%%%%%%%%%%%%%%%%%%%%%%%%%%%%%%%%%%%%%%%%%%%%%%%%%%%%%%%%%%%%%%%%%%%%%%%%%%%%%%%%%%

\begin{acknowledgements}
We would like to express our thanks to the staff of the Effelsberg 100-m telescope and the Parkes 64-m telescope for their assistance with the observations. M.~Ilgner contributed with helpful discussions. C.M. acknowledges support from the Deutsche Forschungsgemeinschaft (DFG) through grant SCHR665/7-1 during part of this study. We thank an anonymous referee and M.~Walmsley for valuable comments and suggestions which helped to improve this paper.
\end{acknowledgements}

\end{document}